\documentclass[11pt,a4paper]{article}
\usepackage{jheppub}
\usepackage{tikz}
\usepackage{xcolor}
\usepackage{mathtools}
\usepackage{verbatim}
\usepackage{adjustbox}
\usepackage{subfigure}
\usepackage[inline,final]{showlabels}
\usetikzlibrary{decorations.pathreplacing}
\usetikzlibrary{decorations.markings}
\usetikzlibrary{calc}
\newcommand{\im}{\mathbf{i}}
\newcommand{\R}{\mathbb{R}}
\newcommand{\Z}{\mathbb{Z}}
\newcommand{\C}{\mathbb{C}}

\renewcommand{\Re}{\mathrm{Re}}
\renewcommand{\Im}{\mathrm{Im}}

\title{On the analytical continuation of lattice Liouville theory}

\author[a]{Xiangyu Cao,}
\author[b]{Raoul Santachiara,}
\author[a,b]{Romain Usciati}
\affiliation[a]{Laboratoire de Physique de l'Ecole Normale Sup\'erieure,\\ ENS, Universit\'e PSL, CNRS, Sorbonne Universit\'e, Universit\'e de Paris, 75005 Paris, France}

\affiliation[b]{Universit\'e Paris-Saclay, CNRS, LPTMS, 91405, Orsay, France}

\emailAdd{xiangyu.cao@phys.ens.fr}

\emailAdd{raoul.santachiara@universite-paris-saclay.fr}

\abstract{The path integral of Liouville theory is well understood only when the central charge $c\in [25, \infty)$.  Here, we study the analytical continuation the lattice Liouville path integral to generic values of $c$, with a particular focus on the vicinity of $c\in (-\infty, 1]$. We show that the $c\in [25, \infty)$ lattice path integral can be continued to one over a new integration cycle of complex field configurations. We give an explicit formula for the new integration cycle in terms of a discrete sum over elementary cycles, which are a direct generalization of the inverse Gamma function contour. Possible statistical interpretations are discussed. We also compare our approach to the one focused on Lefschetz thimbles, by solving a two-site toy model in detail. As the parameter equivalent to $c$ varies from $[25, \infty)$ to $(-\infty, 1]$, we find an infinite number of Stokes walls (where the thimbles undergo topological rearrangements), accumulating at the destination point $c \in (-\infty, 1]$, where the thimbles become equivalent to the elementary cycles. }

\begin{document}
	\maketitle
	
	\section{Introduction}
	Liouville theory is a class of two-dimensional conformal field theories  parametrized by the central charge $c$. The extent to which the theory is understood depends on the value of $c$. In the domain $c\in \C \setminus (-\infty, 1]$, i.e., everywhere in the complex plane except a half-line, the theory is exactly solved by conformal bootstrap. One has a non-compact theory with a continuous spectrum, and correlation functions can be consistently determined from the Dorn-Otto-Zamalodchikov-Zamalodchikov (DOZZ) structure constants~\cite{do94,zz95} and conformal blocks (for a pedagogical account, see \cite{ribault2014conformal}); in particular, correlation functions are analytical in $c \in \C \setminus (-\infty, 1]$. For $c\in [25, \infty)$, the interval relevant to 2D quantum gravity~\cite{polya81}, one can do even better: it is possible to make rigorous sense of the path integral defining the theory, and confirm key aspects of the bootstrap solution from bottom up~\cite{david16,vargas2019,vargas2020}. (The path integral representation underlies also the connection between $c\ge25$ Liouville theory and random energy models~\cite{lft,lft2}.) 
	
	The situation is more confused on the half-line $c \in (-\infty, 1]$. From the bootstrap perspective, the DOZZ structure constants cannot be analytically continued to generic values of $c \in (-\infty, 1] $. Instead, another set of structure constants were proposed, independently by Schomerus~\cite{Schomerus_2003}, Al. Zamalodchikov\cite{Zam05}, and Petkova-Kostov~\cite{kp05}, as an alternative solution to a set of degenerate crossing symmetry equations~\cite{Te95}. These $c \le 1$ structure constants were considered non-physical for a long time, until a number of recent concurrent results put them back on the table. Among these, the most clear-cut came arguably from conformal bootstrap: Ribault and Santachiara~\cite{RiSa015} showed that the $c\le 1$ structure constants define a consistent conformal field theory with a continuous and diagonal spectrum, by numerical checks of the crossing symmetry. However, the bootstrap does not inform of a path integral or microscopic realization of the theory.
	
	The study of $c\le1$ Liouville theory has long been motivated by critical lattice models in two dimensions, such as the $O(n)$ loop models and the related Potts models. Critical exponents in these models can be exactly determined by arguments involving the Liouville path integral~\cite{kondev}. A series of works in the last decade established the statistical interpretation of the $c\le1$ structure constants in terms of three-point correlation functions in Potts~\cite{Delfino_2011,Ziff_2011,PICCO13} and loop models~\cite{IJS,ang2021integrability}. However, exhaustive recent scrutiny of the four-point correlation functions of Potts and loop models~\cite{Ribault:2022qwq,Grans-Samuelsson:2021uor,Jacobsen:2022nxs,He:2020mhb,He:2020rfk,Picco:2016ilr,LykkeJacobsen:2018cbt} has revealed an operator content that is distinct from --- and considerably more involved than --- the aforementioned bootstrap solution~\cite{RiSa015}: Liouville theory and loop models are not as related as one might have believed.
	
	Independently, motivated by applications to holography and four-dimensional gauge theory, Harlow, Maltz and Witten~\cite{Harlow_2011} undertook an extensive semiclassical analysis of Liouville path integral (see Section~\ref{sec:gen} for more detailed discussion). Their results indicate that, to define the Liouville path integral for generic complex values of $c$ (and in particular for $c\in (-\infty, 1]$) requires to understand what the symbol
	$$\int [\mathcal{D} \varphi] $$ 
	actually means, in particular, on which integration cycle the path integral should be carried out. The continuum path integral being a highly nontrivial mathematical object, we shall be more modest and consider lattice regularizations. To obtain the $c\ge 25$ Liouville, the integration cycle and the corresponding lattice regularization are rather straightforward to specify:
	\begin{equation}\label{eq:contourcge25}
		\int [\mathcal{D} \varphi] = \int_{\varphi \in \R} [\mathcal{D} \varphi]  \longrightarrow  \int_{\R^N} \prod_{r=1}^{N} \mathrm{d} \varphi(x_r) \quad \quad (c \ge 25) \,.
	\end{equation}
  Note that this equation is not as innocent as it may appear. It implies in particular that one should integrate over the global shift of the $\varphi$ field (also known as the zero mode), as well as its fluctuations. The importance of the zero mode was noticed early on in the physics literature~\cite{seiberg90,goulianli}, and re-emphasized in the recent rigorous construction of the path integral using probability theory \cite{david16} (see also~\cite{lft} for application to disordered systems). For other values of $c$, and in particular for $c\le1$, a precise prescription analogous to \eqref{eq:contourcge25} is unknown, besides the general expectation that one should integrate over a cycle of middle dimension in the space of complexified field configurations~\cite{Harlow_2011}, i.e., a submanifold of $N$ real dimensions in $\C^N$ in a lattice with $N$ sites. 

  In this work, we show how to analytically continue the \textit{lattice} Liouville path integral from $c\ge25$ to all values of $c \in \C$. In particular, under mild conditions, we provide an explicit integration cycle $\mathcal{C} \subset \C^N$ (in terms of a discrete sum over ``elementary cycles''), which allows to extend the lattice Liouville path integral to parameter regions where the cycle $\R^N$ is no longer viable, including the vicinity of $c\in (-\infty, 1]$. We will argue that the extended lattice path integral we propose has a continuum limit described by the Liouville bootstrap for $c\notin  (-\infty, 1]$. Meanwhile, for $c\le1$, we expect our proposal to display singularities in the continuum limit. Hence, the definition of a continuum path integral for $c\le 1$ Liouville remains an open question. 

  The rest of the paper is organized as follows. 
  \begin{itemize}
      \item  Section~\ref{sec:onesite} reviews the minisuperspace approximation of Liouville theory, which is equivalent to the Liouville path integral on a single lattice site. The resulting Gamma function integral has been extensively discussed (see \cite{Harlow_2011} and references therein). Our main purpose there will be to introduce the notion of elementary cycles which will play a central r\^ole.
      \item  Section~\ref{sec:twosite} studies in detail the toy model of Liouville path integral on two lattice sites. We demonstrate the elementary cycle decomposition in Section~\ref{sec:elementary-twosite}, using a method that will be generalized to general lattices. In Section~\ref{sec:zeromode-twosite} and \ref{sec:CGtwosite} we apply the zero mode approach to the toy model and study the related Coulomb gas integral. Section~\ref{sec:stokes} is devoted to the Stokes phenomena in this toy model, which turn out to be rather involved.
      \item Section~\ref{sec:gen} considers the general lattice Liouville path integral, adapting the methods already exhibited in Section~\ref{sec:twosite}. The main result can be found in Section~\ref{sec:elementary-gen}, equation~\eqref{eq:ZRn-decomp}. For the impatient reader, only Sections~\ref{sec:onesite} and \ref{sec:elementary-twosite} are needed to understand Section \ref{sec:elementary-gen}. 
  \end{itemize}

  

	\section{Minisuperspace approximation}\label{sec:onesite}
	The issue of choosing an integration cycle can be already illustrated in a minisuperspace approximation, where we only integrate over constant field configurations~\cite{Harlow_2011}. Equivalently, we can think of a lattice Liouville theory on a single site. This gives rise to a one-variable integral
	\begin{equation}\label{eq:gamma-integral} 
		\int_{\mathcal{C}} \exp( -n \varphi -  e^{\varphi} )  \mathrm{d} \varphi \,.
	\end{equation}
	This integral is convergent on the $c\ge25$ contour $\mathcal{C} =\R$ if and only if $\Re(n) < 0$ (this is known as a Seiberg bound~\cite{seiberg90} in the context of Liouville), in which case it can be related to the usual gamma function integral by a change of variable $t =  e^{\varphi}$: 
	\begin{equation}
		\int_{\R} \exp( -n \varphi -e^{\varphi} ) = \int_0^\infty t^{-n-1} e^{-t} \mathrm{d} t = \Gamma(-n) \,.  \label{eq:gamma-integral-line}
	\end{equation}
	Thus, the result of the integral can be analytically continued beyond the Seiberg bound to a meromorphic function of $n \in \C$, with poles at $n = 0,1,2,3\dots$.

 \begin{figure}
     \centering
     \includegraphics[scale=0.9]{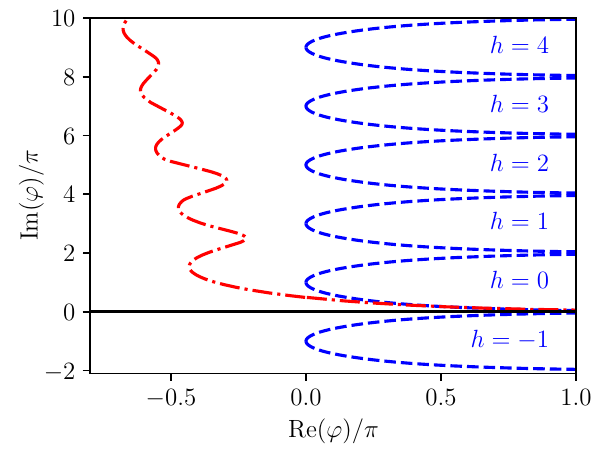}
     \caption{Integration cycles of the Gamma function integral \eqref{eq:gamma-integral} in the complex plane. The integral converges on the black solid line $\varphi \in \R$ when $\Re(n) < 0$. In that case $\R$ is also a Lefschetz thimble. The blue dashed curves are the elementary cycles, indexed by $h \in \Z$. The integral is always convergent on the elementary cycles, and gives essentially an inverse Gamma function~\eqref{eq:gamma-integral-U}. These are equivalent to the thimbles for $\Re(n) > 0$. Finally the red curve is a thimble when $n = e^{-\im \theta}$ for $\theta = \pi /2 - 0.01$, at the vicinity of a Stokes wall.}
     \label{fig:gamma-thimble}
 \end{figure}
	Now, there is a class of cycles, dubbed ``elementary cycles'', on which the integral is convergent for any $n \in \C$. These are indexed by an integer $h\in\Z$, and have a shape of the letter ``U'' rotated by $90$ degrees clockwise. We can describe them as a parametric curve  $\varphi(s), s \in (-\infty, +\infty)$, with the following limiting behavior
	\begin{align}\label{eq:Undef}
		\mathcal{U}_h: \varphi(s)  \longrightarrow \begin{cases} + \infty + 2\pi \im h  & s \to + \infty  \\
			+ \infty + 2\pi \im (h+1) & s \to - \infty 
		\end{cases}  
	\end{align} 
    Such cycles are depicted by blue dashed curves in Figure~\ref{fig:gamma-thimble}.
	By Cauchy's theorem, the precise geometric shape of curve does not affect the integral. In both limits $s\to\pm\infty$, $ -n \varphi -e^{\varphi}  \sim e^{\varphi} \to - \infty $, so the integral \eqref{eq:gamma-integral} converges rapidly, for any value of $n$.  Hence, the integral over $ \mathcal{U}_h $ is an entire function of $n$. In fact one can show that
	\begin{align}\label{eq:gamma-integral-U}
		\int_{\mathcal{U}_h} \exp( -n \varphi -e^{\varphi} ) \mathrm{d}\varphi &= \Gamma(-n) (e^{-2 \pi    h n \im }  - e^{-2 n \pi (h+1) \im  } )
		\\ &=   \frac{2\pi \im   }{\Gamma(1+n)}  \,  e^{- (2h+1) n \pi \im } \,. \label{eq:gamma-integral-U1}
	\end{align}
	To see this, we can restrict to $\Re(n) < 0$ (thanks to analyticity) and deform the contour into the difference between two horizontal lines $\R + 2\pi h \im$ and $\R + 2\pi (h+1) \im $, evaluate the integral using \eqref{eq:gamma-integral-line} and use reflection formula  $\Gamma(-n) \Gamma(1 + n) = \pi / \sin(-\pi n)$ to obtain the expression \eqref{eq:gamma-integral-U1}, which is manifestly analytical in the whole complex plane. 
	
	Comparing \eqref{eq:gamma-integral-line} and \eqref{eq:gamma-integral-U}, we see that the integral on the $c\ge25$ cycle can be written as an infinite sum of the elementary cycles:
	\begin{equation} \label{eq:gamma-decomp}
		\int_\R \omega =
		\begin{dcases} 
			\sum_{h = 0}^{\infty} \int_{\mathcal{U}_h} \omega   & \Im(n) < 0 \\ 
			- \sum_{h = -1}^{-\infty}\int_{\mathcal{U}_h}  \omega  & \Im(n) > 0 
		\end{dcases}   \,,\, 
	\end{equation}
	where $\omega = \exp( -n \varphi -  e^{\varphi} ) \mathrm{d}\varphi$. Note that we must choose the infinite series depending on the sign of $\Im(n)$ as above, since it is the only convergent one, by \eqref{eq:gamma-integral-U1}. On the other hand, the convergence of the series does not depend on the sign of $\Re(n)$, so the right hand side \eqref{eq:gamma-decomp} allows to extend the left hand side beyond its region of convergence. We have here a (rather trivial) example of the following phenomenon: the analytical continuation of the $c\ge25$ lattice path integral \eqref{eq:gamma-integral-line} gives rise to a discrete sum over elementary cycles. The main point of this paper is that this happens to the lattice Liouville theory beyond the minisuperspace approximation. 
	
	The example of Gamma function integral is often discussed~\cite{pasquetti,berry91,boyd,Harlow_2011,witten2011analytic} with an emphasis on the steepest descent contours (also known as Lefschetz thimbles) and the associated Stokes phenomena (see Section~\ref{sec:stokes} below for a definitions and a novel example). When $n < 0$, $\R$ is the steepest descent contour associated with the saddle point $\varphi_* = \ln (-n) \in \R $ of the ``action'' $S(\varphi) = n \varphi + e^{\varphi}$. Since $S'(\varphi)$ is a periodic function in $\varphi$ with period $2\pi\im$, there are a sequence of saddle points (corresponding to different branches of $\ln(-n)$), and their thimble is always a horinzontal line. As we move $n$ around in $\C$, the steepest descent contours smoothly deform, as long as $\Re(n) < 0$.  When $n$ hits the imaginary axis, the steepest descent contours undergo a topological rearrangement known as the Stokes phenomenon, such that when $\Re(n) > 0$, the thimbles all become equivalent to the elementary cycles $ \mathcal{U}_h $. In order to analytically continue the $c\ge25$ integral, one must also switch abruptly the integration contour from a single thimble to a sum of thimbles as in \eqref{eq:gamma-decomp} to compensate for the Stokes phenomenon. This can be visualized in Figure~\ref{fig:gamma-thimble}, where the red curve is the steepest descent contour just before $n$ hits the Stokes wall. With a larger number of degrees of freedom, characterizing thimbles and the Stokes phenomena can become highly involved~\cite{Alexandru:2020wrj}. Our point of view here is that this difficult task can be avoided by focusing instead on the elementary cycles. 

	\section{Liouville theory on two sites}\label{sec:twosite}
	To further compare the approaches based on thimbles and elementary cycles, it is helpful to consider the simplest toy model of Liouville path integral that involves a ``kinetic term'', which connects two lattice sites. This toy model is simple enough so that the Lefschetz thimbles and the Stokes phenomena can be reasonably understood. They turn out to be considerably more involved than the one-site example above, whereas the elementary-cycle approach yields a more transparent picture.
	
	We consider the integral 
	\begin{equation}\label{eq:twosites}
		Z({\mathcal{C}}) = \int_{\mathcal{C}} \exp\left(- S(\varphi_1, \varphi_2) \right) \mathrm{d} \varphi_1 \mathrm{d} \varphi_2 
	\end{equation}
	where the action 
	\begin{equation} \label{eq:Stwosite}
		S(\varphi_1, \varphi_2) =  \frac1{2b^2} (\varphi_1 - \varphi_2)^2 + \sum_{i=1,2} \left(e^{\varphi_i} - \frac{\eta_i}{b^2} \varphi_i \right)  
	\end{equation}
	depends on the parameters $b$ and $\eta_i, i=1,2$. We can associate $b$ with the central charge $c$ in a way such that 
	\begin{equation} \label{eq:cbrelation}
		\begin{cases}
			c \ge 25  & b \in \R \,,\, b^2 > 0 \\
			c \le 1   & b \in \im \R \,,\, b^2 < 0 \,.
		\end{cases} \,.
	\end{equation}
	Indeed, the kinetic term in the action has the usual (``space-like'') sign when $b^2 > 0$, and the ``wrong'' (``time-like'') sign as $b^2 < 0$. Of course, in the case of a two sites model,  the notion of conformal symmetry, and therefore of central charge, is meaningless. However,  changing $c$  from $[25,\infty)$ to $(-\infty,1]$ does correspond to flipping the sign of the kinetic term in the full-fledged Liouville path integral, see \eqref{eq:cbrelation-real} below. Similarly, anticipating \eqref{eq:aQandeta} below, we associate $\eta_i$ with operator insertions in Liouville theory. 
	
	When $b^2 > 0$, the path integral on $\R^2$ converges provided $\eta_i$ satisfy the Seiberg bound:
	\begin{equation}
		\sum_{i} \Re(\eta_i) > 0\,.
	\end{equation} 
	We will be interested in the continuation of the $c\ge 25$ path integral to the vicinity of $c \in (-\infty,1]$. This corresponds to rotating $b^2$ from the positive real axis to the negative one. For definitiveness, we choose to do it \textit{clock-wise}:
	\begin{equation}\label{eq:theta}
		b^2 = |b^2| e^{-\im \theta} \,,\, \theta = 0 \to \pi \,.
	\end{equation}
	For simplicity, we fix $\eta_i$ to be equal and positive: 
	\begin{equation}
		\eta_1 = \eta_2 = \eta > 0 \,.
	\end{equation}
   This restriction will be relaxed in a more general setting in Section~\ref{sec:gen} below.
	
	\subsection{Elementary cycle decomposition}\label{sec:elementary-twosite}
\begin{figure}
     \centering
     \includegraphics[scale=0.8]{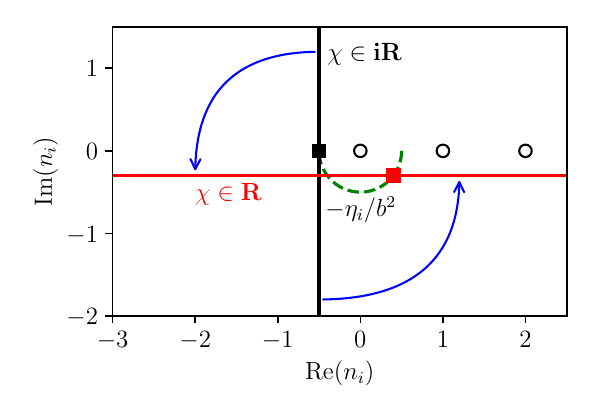}
     \caption{The possible values of $n_i$~\eqref{eq:n1n2} during the deformation of the contour of $\chi$. The latter is $\im \R$ when $b^2 > 0$, so $n_i$ belongs to a vertical line shifted by $-\eta_i / b^2 < 0$ (indicated by the black square). As $b^2 = |b^2| e^{-\im \theta}$ rotates clockwise ($\theta $ increases from $0$ to $\pi$), the $\chi$ contour rotates anti-clockwise towards $\R$. Meanwhile $-\eta_i/b^2$ follows the green dashed half-circle below the real axis, and arrives at the red square as $\theta = \pi - \epsilon$. Then, $n_i$ belongs to the horizontal red line, and always satisfies $\Im (n_i) < 0$ (which is important for the elementary cycle decomposition, see above \eqref{eq:twosite-decomp} and also \eqref{eq:gamma-decomp}) . We observe that $n_i$ is away from the poles of $\Gamma(-n_i)$ (empty dots) throughout the contour deformation process.   }
     \label{fig:ni-contour}
 \end{figure}

	A useful way to analyze \eqref{eq:twosites} is to decouple the kinetic term in \eqref{eq:Stwosite} by introducing an auxiliary variable $\chi$:
	\begin{align}
		&    \exp\left(- S(\varphi_1, \varphi_2) \right) = \int_\chi \frac{e^{b^2 \chi^2 / 2}\mathrm{d}\chi}{ \im \sqrt{2\pi /b^2} }  \prod_{i=1,2} \exp\left( - n_i \varphi_i - e^{\varphi_i} \right) \label{eq:decouple2}  \\ 
		& n_1 = - \frac{\eta}{b^2} + \chi \,,\, n_2 = -  \frac{\eta}{b^2} - \chi   \label{eq:n1n2}
	\end{align}
	The contour of $\chi$ can be chosen arbitrarily, so long as the Gaussian integral $ \int_\chi e^{ b^2 \chi^2 / 2} $ converges. For example, a vertical (horizontal) contour is admissible whenever $\Re(b^2) > 0$ ($<0$, respectively). For any fixed $\chi$, the action~\eqref{eq:decouple2} factorizes into a one-site action~\eqref{eq:gamma-integral} for each $\varphi_i$, so we can use the results of previous section to integrate out $\varphi_i$ first. In particular, we know that the  integral over the $c\ge 25$ contour, whenever that converges, gives the following 
	\begin{equation}
		Z(\R^2) = \int_\chi \frac{e^{b^2 \chi^2 / 2}\mathrm{d}\chi}{ \im \sqrt{2\pi /b^2} }\prod_{i=1,2} \Gamma(-n_i)  \,. \label{eq:ZR2-int}
	\end{equation}
	It follows that the left hand side can be analytically continued beyond where the $\varphi_i$-integrals converge, as long as one can smoothly deform the $\chi$-contour to keep the $\chi$ integral convergent, while avoiding the poles of the Gamma functions. One can check that this can be done for all $0< \theta < \pi$ (recall $b^2 = |b^2| e^{-\im \theta}$), by rotating the contour of $\chi$ to $e^{\im(\pi /2+ \theta)} \R $, see Figure~\ref{fig:ni-contour}.

	We now consider the elementary cycles in the two-site model. They are indexed by $(h_1, h_2) \in \Z^2$, and are defined as a product of the one-site elementary cycles
	\begin{equation}\label{eq:Udef-twosite}
		\mathcal{U}_{h_1,h_2} := \prod_{i=1,2} \mathcal{U}_{h_i}(\varphi_i) \,,
	\end{equation}
	where $\mathcal{U}_h$ is the one-site elementary cycle defined in \eqref{eq:Undef}. It is not hard to convince oneself that the path integral is convergent in any $ \mathcal{U}_{h_1,h_2}$ for any value of $b^2$ and $(\eta_i)$.
   Indeed, as $\Re (\varphi_i)\to \infty$ and $\Im(\varphi_i)\to 2\pi h_i$, $h_i\in\Z$, that is, along the elementary cycle $U_{h_i}$, the action \eqref{eq:Stwosite} is dominated by $\sum_i e^{\varphi_i}$, and its real part diverges rapidly to $+\infty$. Using the decoupled action \eqref{eq:decouple2} and \eqref{eq:gamma-integral-U1}, we can find a representation of the path integral on an elementary cycle in terms of an integral over $\chi$:
	\begin{equation}\label{eq:ZU2-int}
		Z(\mathcal{U}_{h_1,h_2}) = 
		\int_\chi \frac{e^{b^2 \chi^2 / 2}\mathrm{d}\chi}{ \im \sqrt{2\pi /b^2} } 
		\prod_{i=1,2} \frac{2\pi \im e^{-(2h_i+1) n_i \pi \im}}{\Gamma(1+n_i)}
	\end{equation}
	
    We turn to discuss the decomposition of $\Z(\R^2)$, or rather the analytical continuation thereof, into a sum of elementary cycles. \emph{For any fixed $\chi$}, the one-site decomposition formula \eqref{eq:gamma-decomp} allows us to write the $\varphi_i$-integral
	$$ \int_{\R^2}   \prod_{i=1,2} \exp\left( - n_i \varphi_i - e^{\varphi_i} \right) \mathrm{d} \varphi_i $$
	as a sum over elementary cycles. However, note that the sum depends on the sign of $\Im(n_{1,2})$, and $n_{i} = -\eta_i/b^2 \pm \chi$ depend on $\chi$. Therefore, we can decompose the whole path integral into a sum of elementary cycles \textit{if} we can choose a contour of $\chi$ such that $\Im(n_i)$ does not change sign. Such a contour must be horizontal, i.e., $\chi \in \R + \im c $. A horizontal contour integral is convergent if $\Re(b^2) < 0$, which is satisfied for $\theta\in (\pi/2,\pi)$. Since $\Im(b^2) < 0$ for all $\theta \in (0, \pi)$, we can choose the contour $\chi \in \R$ and have $\Im(n_i) <  0$ throughout, see Figure~\ref{fig:ni-contour}. As a result, by \eqref{eq:gamma-decomp}, we have 
	\begin{equation}\label{eq:twosite-decomp}
		Z({\R^2}) =  \sum_{h_1 = 0}^\infty  \sum_{h_2 = 0}^\infty Z({\mathcal{U}_{h_1,h_2}}) \,,\, \theta \in (\pi/2, \pi) \,.
	\end{equation}
	This formula is the generalization of the one-site decomposition formula \eqref{eq:gamma-decomp}. Like the latter, \eqref{eq:twosite-decomp} depends on our choice \eqref{eq:theta} of continuing $b$ clockwise. Otherwise, we would have a sum over $h_1 < 0, h_2< 0$. Note that when \eqref{eq:twosite-decomp} is applicable, the left hand side is not convergent and should be understood as an analytical continuation, which can be performed using \eqref{eq:ZR2-int}. Then, we can use \eqref{eq:twosite-decomp} to examine the analytical properties of $Z(\R^2)$ near the $ c \le 1$ line. Observe that the action behaves simply under a global shift in the imaginary direction, $S(\varphi_1 + 2\pi \im h, \varphi_2 + 2\pi \im h) = S(\varphi_1, \varphi_2) -2\pi \im h \sum_{i} \eta_i  / b^2$, where $h\in \mathbb{Z}$. Since the elementary cycles are related by a translation in the imaginary direction~\eqref{eq:Undef}, we have 
	\begin{equation}
		Z({\mathcal{U}_{h_1,h_2}}) = Z({\mathcal{U}_{h_1 + h,h_2 + h}}) e^{2\pi\im h \sum_i\;\eta_i / b^2} \,.
	\end{equation}
	This, we can parametrize the terms in \eqref{eq:twosite-decomp} as $(h_1, h_2) = (h_1' + h_0, h_2'+h_0)$ with $h_0 = \min (h_1, h_2)$ and $\min (h_1', h_2') = 0$, and transform the sum as follows: 
	\begin{align}
		Z({\R^2}) = \widehat{Z} \sum_{h_0=0}^{\infty} {e^{4\pi\im h_0 \eta / b^2}}  = \frac{\widehat{Z}}{1-e^{4\pi\im \eta / b^2}}   \,,\, \theta > \pi /2 \,, \label{eq:ZR2andZhat}
	\end{align}
	where 
	\begin{equation}\label{eq:Zhat}
		\widehat{Z} := \sum_{\min(h_1, h_2) = 0} Z(\mathcal{U}_{h_1,h_2}) = 
		Z(\mathcal{U}_{0,0}) + \sum_{h=1}^{\infty} ( Z(\mathcal{U}_{h,0}) + 
		Z(\mathcal{U}_{0,h}))  \,.
	\end{equation}
	$\widehat{Z}$ is a convergent sum when $\Re(b^2) < 0$. Indeed, for $h$ large, $\Im(\varphi_1 - \varphi_2) \sim  \pm 2\pi \im h$ for all $(\varphi_i) \in \mathcal{U}_{h,0}$ or $ \mathcal{U}_{0,h}$, the integral over these cycles are suppressed by the ``time-like'' kinetic term in \eqref{eq:Stwosite} (the contribution from regions where some $\Re(\varphi_i)$ becomes large and positive is further suppressed by $e^{\varphi_i}$, which dominates over the time-like kinetic term). Therefore, \eqref{eq:ZR2andZhat} isolates the singularities of $Z(\R^2)$ as we approach $b^2 < 0$: we have a series of poles at 
	\begin{equation} 
		b^2 = - 2\eta / n \,,\, n = 1, 2, 3,\dots \,.  
	\end{equation}
	
	The upshot of the above analysis is the following.  The $c\ge25$ path integral can be continued and decomposed into a sum over elementary cycles \eqref{eq:twosite-decomp} with nonnegative indices $h_i \ge 0$. The singularities as we approach the $c\le1$ line are due to the sum over the ``zero mode'', i.e., global shifts of the indices, and can be removed by considering a restricted sum over elementary cycles with zero minimum index, $\min h_i = 0$.
	
	\subsection{The zero mode approach}\label{sec:zeromode-twosite}
	\newcommand{\phid}{{\tilde{\varphi}}}
	The goal of this section and the next one is to understand the above result in light of the zero mode approach, which is an important step in the analysis of the $c\ge25$ Liouville path integral~\cite{goulianli,david16}. The idea is to write the Liouville field as a sum of the zero mode $\varphi_0$ and the fluctuating ones $\phid$~\cite{goulianli,david16}. In our toy model, there is only one fluctuating mode, and we can write:
	\begin{equation} \label{eq:twositezeromode-def}
		\varphi_0 = \frac12 (\varphi_1 + \varphi_2)  \,,\,  \phid = \varphi_1 - \varphi_2  \; \Leftrightarrow \;
		\varphi_1 = \varphi_0 + \frac{\phid}2 \,,\, \varphi_2 = \varphi_0 - \frac{\phid}2  \,.
	\end{equation}
	In terms of these variables the action is 
	\begin{equation}\label{eq:S_phi0phid}
		S = \frac{\phid}{2b^2} + 2 \cosh\left(\frac{\phid}2\right) e^{\varphi_0} - \frac{ 2 \eta \varphi_0}{b^2}
	\end{equation}
	Now, for any fixed $\phid$, upon a shift, 
	\begin{equation} \label{eq:zeromode_shift}
		\varphi_0 = \varphi - \ln \left(2 \cosh\left( \frac{\phid}2 \right) \right) \,,
	\end{equation}  
	we have a one-site action~\eqref{eq:gamma-integral} for $\varphi$, with $n = -2\eta/b^2$. Therefore, to calculate $Z(\R^2)$, we can integrate first over $\varphi_0$ over $\R$, which results in an integral over the fluctuating mode:
	\begin{equation}
		Z(\R^2) = 
		\Gamma\left(\frac{2\eta}{b^2} \right) \int_{\R} \mathrm{d} \phid \, \exp( - S_{\text{eff}}(\phid) ) \,,\,  \label{eq:ZR2Seff}
	\end{equation}
	with an effective action:
	\begin{equation}
		S_{\text{eff}}(\phid) = \frac{\phid^2}{2b^2} + \frac{2\eta}{b^2}\ln \left(2 \cosh\left( \frac{\phid}2 \right) \right) \,, \label{eq:Seff}
	\end{equation}
	where we choose the branch of the log so that $\ln(2\cosh(\phid/2)) \in \R$ for all $\phid \in\R$. Eq.~\eqref{eq:ZR2Seff} is another integral representation of $Z(\R^2)$, ``dual'' to \eqref{eq:ZR2-int}. As it is written, it is convergent if and only if $\Re(b^2) > 0$. To continue it beyond, we can deform the contour of $\phid$ smoothly without crossing the branch cut singularities of $S_{\text{eff}}(\phid)$. As $b^2 = |b^2|e^{-\im\theta}$ rotates clockwise from $\R_+$ to $\R_-$~\eqref{eq:theta}, we can rotate the $\phid$ contour clockwise, from $\R$ to $(\im-\epsilon) \R$; in fact, the latter contour is convergent for all $\theta > \pi /2$:
	\begin{equation} \label{eq:ZR2-zeromode}
		Z(\R^2) = 
		\Gamma\left(\frac{2\eta}{b^2} \right) \int_{(\im - \epsilon)\R} \mathrm{d} \phid \, \exp( - S_{\text{eff}}(\phid) ) \,,\, \theta > \pi /2 \,.
	\end{equation}
	Comparing that with \eqref{eq:ZR2andZhat}, and using the reflection formula, we find
	\begin{equation}\label{eq:Zhat-zeromode}
		\widehat{Z} = \frac{2\pi \im e^{2\pi \im \eta/b^2}}{\Gamma\left(1 -\frac{2\eta}{b^2} \right)} \int_{(\im - \epsilon)\R} \mathrm{d} \phid \, \exp( - S_{\text{eff}}(\phid) ) \,.
	\end{equation}
	The above two formulas tell us that both $Z(\R^2)$ (continued to near $c\le 1$) and $\widehat{Z}$ are a product of an integral over a shifted zero mode $\varphi$, and an integral over the fluctuating mode. Their difference lies in the choice of the zero-mode cycle: $Z(\R^2)$ integrates over $\varphi \in \R$, and $\widehat{Z}$ over an elementary cycle $\mathcal{U}_{0}$, see \eqref{eq:gamma-integral-U1}. 
\begin{figure}
    \centering
    \includegraphics[scale = 0.8]{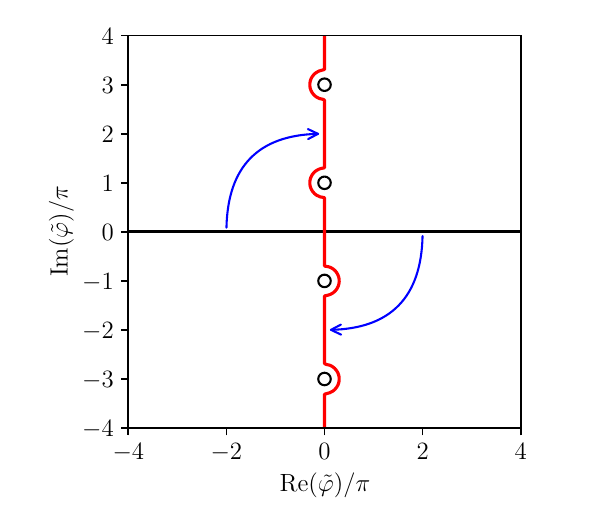}
    \caption{The deformation of the contour of the fluctuation mode $\phid$, from $\R$ (black line) to $(\im-\epsilon) \R$ (red curve), in order to analytically continue \eqref{eq:ZR2-zeromode} as $b^2 = |b^2| e^{-\im\theta}$ and $\theta$ increases from $0$ to $\pi$. The empty dots are the branch cut singularities of $S_{\text{eff}}(\phid)$. }
    \label{fig:varphi_contour}
\end{figure}

	The fluctuating mode integral is closely related to the Coulomb gas formalism~\cite{dofa_npb84}. This relation is the most transparent when $\theta = \pi$, so that we can write $b = \im \beta$, and $n = -2\eta / b^2 = 0,1,2,3\dots$; then a change of variable $\phid = \im \beta y$ gives
	\begin{equation}
		\widehat{Z} = \frac{(-1)^n 2\pi\beta}{n!} 
		\int_{\R} e^{-y^2/2} (e^{ \im \beta y} + e^{-\im \beta y})^n  \mathrm{d} y
	\end{equation}
	As the integrand becomes singled valued, the contour need no longer be tilted, and $y$ can be viewed as parametrizing a Gaussian field $(y, -y)$ with zero mean on the two site model. The observable can be expanded as a sum over configurations of $n$ charges on two sites:
 \begin{equation}
    \frac1{n!}(e^{ \im \beta y} + e^{-\im \beta y})^n = 
	\sum_{n_1 + n_2 = n} \frac{e^{\im \beta y n_1 }}{n_1!} 
	\frac{e^{- \im \beta y n_2}}{n_2!} 
 \end{equation}  
 In a lattice model with many sites, one would sum/integrate over the positions of $n$ charges: this is known as a Coulomb gas integral~\footnote{In this work, since we work only with discrete lattices, we do not integrate over charge positions. Instead, we use the word ``integral'' in ``Coulomb gas integral'' to refer to that over the fluctuating mode $\phid$. }. The Liouville path integral for generic values of $n$ is an extension of the Coulomb gas approach. That requires a contour prescription for the multi-valued function $(e^{ \im \beta y} + e^{-\im \beta y})^n$. The analytical continuation of $b^2$ from $\R_+$ to $\R_-$ provides such a prescription. 

	\subsection{Elementary cycles of the Coulomb gas integral}\label{sec:CGtwosite}
	How are the results of the zero approach \eqref{eq:ZR2-zeromode} and \eqref{eq:Zhat-zeromode} related to the elementary cycle decompositions \eqref{eq:twosite-decomp} and \eqref{eq:Zhat}? To address this question, we examine the Coulomb gas integral on a general cycle:
	\begin{equation}\label{eq:CGcycle}
		\widetilde{Z}(\mathcal{C}):= \int_{\mathcal{C}} \mathrm{d} \phid \, \exp( - S_{\text{eff}}(\phid) ) \,.
	\end{equation}
	Throughout this section,  we shall restrict ourselves to the near time-like region $\Re(b^2) < 0$. Since $\eta > 0$, the real part of the effective action goes to $+\infty$ at the zeros of $\cosh(\phid/2)$:
	\begin{equation}
		\Re(S_{\text{eff}}) \to +\infty \,,\, \phid \to (2\tilde{h} + 1)\pi \im \,,\, \tilde{h} \in \Z \,.
	\end{equation}
	Therefore, for any $\tilde{h}$, the interval
	\begin{equation}\label{eq:Idef0}
		\mathcal{I}_{\tilde{h}} = ((2 \tilde{h} - 1)\pi \im, (2 \tilde{h} + 1)\pi \im )
	\end{equation} 
	is a closed integration cycle on which \eqref{eq:CGcycle} converges. However, due to the multi-valuedness of \eqref{eq:Seff},  the cycles should be defined on a cover space of  $\C - \{(2\tilde{h} + 1)\pi \im: \tilde{h} \in \Z \}$ on which $S_{\text{eff}}$ is single-valued. In practice, we can specify the deck by the value of the log, say at the mid-point of $  \mathcal{I}_{\tilde{h}} $:
	\begin{equation}\label{eq:Idef1}
		\mathcal{I}_{\tilde{h}}^{h_0} =  ((2 \tilde{h} - 1)\pi \im, (2 \tilde{h} + 1)\pi \im ) \,,\, \ln\cosh(\im \pi \tilde{h} ) =  - 2 \pi h_0  \im  \,.
	\end{equation}
	Here, $h_0 \in \Z / 2$ such that 
	\begin{equation} \label{eq:h0htilde}
		h_1 = h_0 + \frac{\tilde{h}}2  \,,\, h_2 = h_0 -  \frac{\tilde{h}}2 
	\end{equation}
	are both integers (compare to \eqref{eq:twositezeromode-def}). Hence, we have again a lattice of cycles
	\begin{equation}\label{eq:Idef2}
		\mathcal{I}_{h_1, h_2} := \mathcal{I}_{\tilde{h}}^{h_0 }  \,,\, \tilde{h} = h_1 - h_2 \,,\, h_0 = \frac12 (h_1 + h_2)
	\end{equation}
	indexed by $(h_1, h_2) \in \Z^2$, which we call the elementary cycles of the Coulomb gas integrand.  Now, it is not hard to verify that the Coulomb gas integral in \eqref{eq:Zhat-zeromode} admits the following decomposition 
	\begin{align}
		\int_{(\im - \epsilon)\R} \mathrm{d} \phid \, \exp( - S_{\text{eff}}(\phid) )  =& 
		\widetilde{Z}(\mathcal{I}_{0}^{0}) + \sum_{\tilde{h} = 1}^\infty     
		\left( \widetilde{Z}\left(\mathcal{I}^{\tilde{h} / 2}_{\tilde{h}} \right) + 
		\widetilde{Z}\left(\mathcal{I}^{\tilde{h} / 2}_{-\tilde{h}}\right) \right)
		\nonumber \\
		=& \sum_{\min(h_1,h_2) = 0} 
		\widetilde{Z}(\mathcal{I}_{h_1, h_2}) \,. \label{eq:CG-decomp}
	\end{align}
	Note that $h_0$ increases as $\tilde{h} \to \pm  \infty$ precisely because, as we go from $0$ to $\pm (\im -\epsilon) \infty$ on the contour of \eqref{eq:Zhat-zeromode}, we always half-encircle the zeros of $\cosh$ clock-wise, in both directions; see Figure~\ref{fig:varphi_contour}. Eq.~\eqref{eq:CG-decomp} is clearly reminiscent of \eqref{eq:Zhat}. This is no coincidence. In fact, we can prove the following relation between the elementary cycles (see Appendix \ref{sec:proofZandZCG}): 
	\begin{equation}\label{eq:ZandZCG}
		Z(\mathcal{U}_{h_1, h_2}) = \frac{2\pi \im e^{2\pi \im \eta/b^2}}{\Gamma\left(1 -\frac{2\eta}{b^2} \right)} \widetilde{Z}(\mathcal{I}_{h_1, h_2}) \,.
	\end{equation}
	Combining equations \eqref{eq:Zhat-zeromode}, \eqref{eq:CG-decomp} and \eqref{eq:ZandZCG}, we recover the elementary cycle decomposition \eqref{eq:Zhat}. 
	
	Let us summarize what we learned so far. The elementary cycles of the Coulomb gas integral are in one-to-one correspondence with those of the two-site path integral. The continuation of the Coulomb gas integral to the time-like region is an infinite sum of elementary cycles. This sum is qualitatively distinct from that of the one-site decomposition (or the Gamma function integral) as no two cycles are simply related. 
	
	Since essential feature of the two-site Liouville integral can be captured by the one-variable Coulomb gas integral, we can use the latter as a proxy to understand the Stoked phenomena of the (two-site) Liouville action as we continue it from $c\ge25$ to $c\le1$. This is the subject of the next section. 
	
	\subsection{Stokes phenomema of the Coulomb gas integral}\label{sec:stokes}
	
	
	In this section we study the saddle points, thimbles and Stokes phenomena of the Coulomb gas effective action, which we recall here: 
	\begin{equation}\label{eq:Seff-recall}
		S_{\text{eff}}(\phid) = \frac{\phid^2}{2b^2} + \frac{2\eta}{b^2}\ln \left(2 \cosh\left( \frac{\phid}2 \right) \right) 
	\end{equation}
	The saddle points are by definition where the derivative of the action vanishes, $ \partial_{\phid}S_{\text{eff}} = 0$. Here, they are  independent of $b$, and can be found as the solutions to a familiar transcendental equation:
	\begin{equation} \label{eq:phid-trans-eq}
		\phid = \im y \,,\, 
		y = -  \eta \tan\left(\frac{y}2 \right) \,.
	\end{equation}
	As $\eta > 0$, there is exactly one saddle point in each interval $  \mathcal{I}_{\tilde{h}}$~\eqref{eq:Idef0} connecting two neighboring zeros of the $\cosh$; so we shall call that saddle point $\phid_{\tilde{h}} = \im y_{\tilde{h}}$. Strictly speaking, because of the multivaluedness of $S_{\text{eff}}$, each saddle point $\phid_{\tilde{h}}$ is the projection of a series of identical copies $\phid^{h_0}_{\tilde{h}}$ where $h_0 \in \Z/2$ is such that $h_0 \pm \tilde{h}/2$ are integers, see \eqref{eq:h0htilde} above. The same can be said about their thimbles. However, to keep the notations light we shall suppress the dependence on $h_0$ unless when it is absolutely necessary to display it.
	
	The thimble (usually known as the steepest descent contour) of $\phid_{\tilde{h}}$, which we denote as $\mathcal{T}_{\tilde{h}}$, is defined by the solutions to the ``upward flow'' equation~\cite{Alexandru:2020wrj,witten2010new} that emanate from the saddle point:
	\begin{equation}
		\overline{ \partial_s \phid(s) } = \partial_{\phid} S_{\text{eff}}(\phid(s))  \,,\,  \lim_{s\to-\infty}\phid =  \phid_{\tilde{h}} \,.
	\end{equation}
	It is well-known that, as $\phid$ flows along the thimble and away from the saddle point, the real part of the action increases while the imaginary part is a constant:
	\begin{equation}
		\Re \left[ \frac{\mathrm{d}  S_{\text{eff}}}{\mathrm{d}s}\right]  > 0 \,,\, \Im  \left[\frac{\mathrm{d}  S_{\text{eff}}}{\mathrm{d}s} \right] = 0 \,.
	\end{equation}
	As a result, as $s \to +\infty$, one of the followings happen: (i) $S_{\text{eff}} \to \infty$; this implies that $\phid \to\infty$  or $\phid$ tends to a zero of $\cosh$ in \eqref{eq:Seff-recall} (the latter can happen only if $\mathrm{Re}(b^2)<0$, since $\eta> 0$); (ii) the thimble converges to a saddle point (which can be the same as $ \phid_{\tilde{h}}$): $\phid(s) \to \phid_{\tilde{h}'}$.  The latter can happen only if the action has the same imaginary part on both saddles: 
	\begin{equation} \Im (S_{\text{eff}}(\phid_{\tilde{h}'})) - \Im (S_{\text{eff}}(\phid_{\tilde{h}})) = 0  \,.	 \label{eq:ImSeff}    
	\end{equation} 
	As the left hand side of this equation smoothly depends on the parameters of the action, the possibility (ii) cannot happen generically and only does so when the parameters are fine-tuned, i.e., when they hit a ``Stokes wall''. As the parameters vary across a Stokes wall, the topology of the thimbles goes through an abrupt transition known as a Stokes phenomenon. 
	
	\begin{figure}
		\centering
		\includegraphics[width=\textwidth]{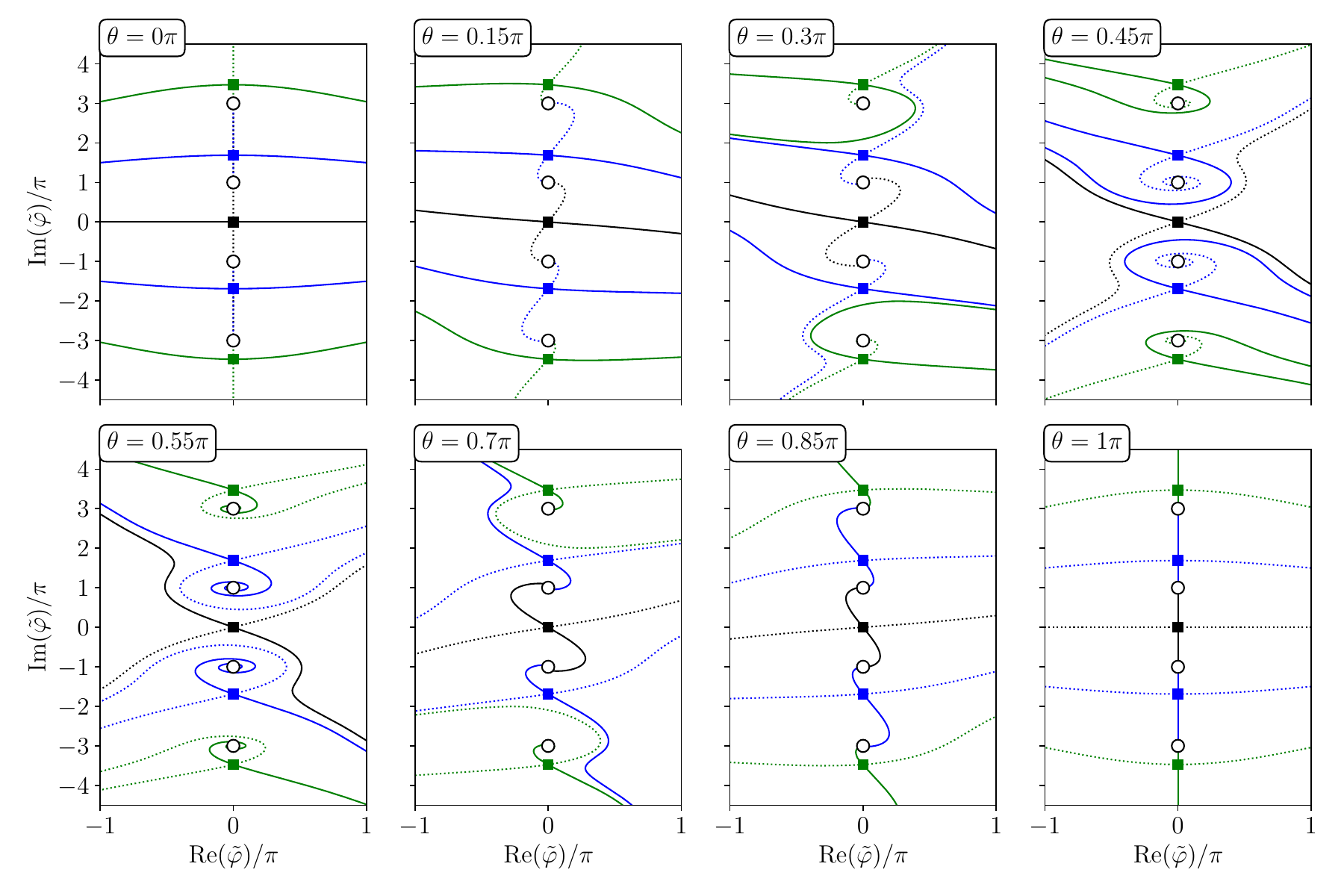}
		\caption{Lefschetz thimbles and anti-thimbles of the one-variable effective action \eqref{eq:Seff-recall} for various values of $b^2 = |b^2| e^{-\im \theta}$, and $\eta = 10$. The solid squares are the saddle points. The empty circles are the branch cut singularities of the log in the action. The solid curves are the thimbles (along which the real part of the action tends to $+\infty$; also known as steepest descent contours); the dashed curves are the anti-thimbles (along which the real part of the action tends to $-\infty$). See Figure~\ref{fig:ah} for further discussion on the behavior of the thimbles, in particular, whether they are equivalent to elementary cycles.  }
		\label{fig:onevar}
	\end{figure}
	These generalities being reviewed, let us work out the explicit example of the Coulomb gas effective action. As before, we let $\eta > 0$ be fixed and rotate $b^2$ from $\R_+$ to $\R_-$ clockwise: $b^2= |b^2|e^{-\im \theta}, \theta \in (0, \pi)$. A panorama of what awaits our analysis is shown in Figure~\ref{fig:onevar}, where we plot the thimbles emanating from a few saddle points close to $\phid=0$. We also plotted the anti-thimbles, which are defined as the thimbles of $-S_{\text{eff}}$. This reveals a nice symmetry: the thimbles at $\theta$ and the anti-thimble $\theta \to \pi - \theta$ are identical (modulo a complex conjugate). 
	
	Figure~\ref{fig:onevar} shows a variety of Stokes phenomena. To orient ourselves, let us first follow the thimble $\mathcal{T}_0$ emanating from the saddle point $0$ (the black solid curve in Figure~\ref{fig:onevar}). At $\theta = 0$, it is simply $\R$, the $c\ge 25$ contour of the fluctuating mode. As $\theta$ increases, at first, the thimble of $0$ deforms smoothly and rotates clockwise roughly speaking. However, after a series of Stokes phenomena (that we will study more carefully below), as we approach $\theta = \pi$, we observe that all the thimbles become bounded, and connect neighboring zeros of $\cosh$. In other words, they become equivalent to the elementary cycles \eqref{eq:Idef1}:
	\begin{equation}\label{eq:IisT}
		\mathcal{T}_{\tilde{h}}^{h_0} \to \mathcal{I}_{\tilde{h}}^{h_0} \,,\, \theta  \to  \pi \,.
	\end{equation}
	As a result, when $c\le1$, the Coulomb gas contour $(\im -\epsilon)\R$ is an infinite sum of thimbles:
	\begin{equation}\label{eq:decom-thimble}
		(\im -\epsilon)\R \to \sum_{\tilde{h} \in \Z} \mathcal{T}_{\tilde{h}}^{|\tilde{h}/2|} \,,\, \theta \to \pi,
	\end{equation}
	similar to \eqref{eq:CG-decomp} above. 

     \begin{figure}
         \centering
         \includegraphics[scale=0.7]{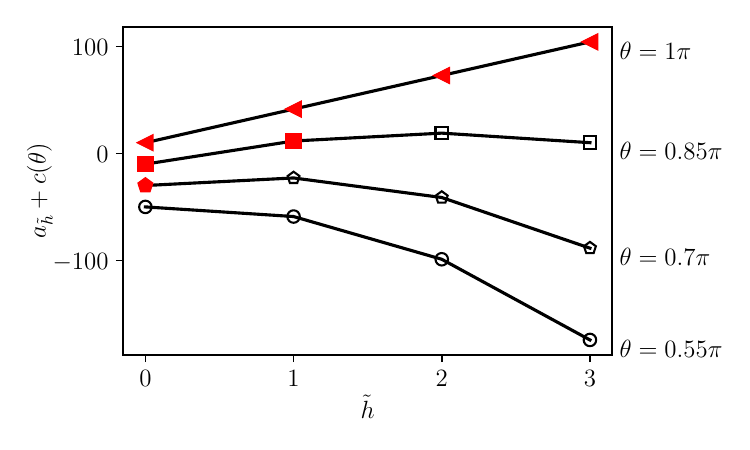}
         \caption{A plot of the first few $a_{\tilde{h}}$'s (the imaginary part of the effective action at the $\tilde{h}$-th saddle point) with the same parameters as in Figure~\ref{fig:onevar}: $\eta = 10$, $b^2 = |b^2| e^{-\im \theta}$, $\theta = 0.55\pi, 0.7 \pi, 0.85 \pi, \pi$. The filled markers indicate the cases where $ a_{\tilde{h}} \le  a_{\tilde{h}+1} $. This implies~\eqref{eq:TisIcondition} that the thimble $\mathcal{T}_{\tilde{h}} = \mathcal{I}_{\tilde{h}} $ is equivalent to the elementary cycle. Compare with Figure~\ref{fig:onevar}. The data for different $\theta$ are shifted for visibility. }
         \label{fig:ah}
     \end{figure}
	So far, eq.~\eqref{eq:IisT} and \eqref{eq:decom-thimble} are conjectures that appear reasonable upon inspecting Figure~\ref{fig:onevar}. We will now show that they are true, but involve a subtlety that is not visible in the figure. To reveal that we shall examine in more detail the Stokes phenomena taking place in $\theta \in (\pi/2, \pi)$. Indeed, they are the ones that are relevant to the decomposition of the Coulomb gas integral (it is apparent in Figure~\ref{fig:onevar} that the thimble $\mathcal{T}_{0}$ is still equivalent to $e^{-\im \theta/2}\R$ when $\theta = \pi/2+\epsilon$). To do this, by \eqref{eq:ImSeff}, it is useful to calculate the imaginary part of the action at the critical points: 
	\begin{equation}\label{eq:ImSeff-exp}
		a_{\tilde{h}} := \Im( S_{\text{eff}} (\phid_{\tilde{h}}^{\tilde{h}/2}  )) = 
		- \pi \eta \tilde{h} \cos(\theta) - \frac{y_{\tilde{h}}^2}{2} \sin(\theta) \,,\, \tilde{h} \ge 0 \,.
	\end{equation}
 See Figure~\ref{fig:ah} for a plot.
	Here and below, we shall restrict to $\tilde{h} \ge 0$ for convenience; the analysis for $\tilde{h} \le 0$ is identical by symmetry. 
	It is not hard to convince ourselves that the relation between $a_{\tilde{h}}$ and the thimbles is the following: the thimble is equivalent to the elementary cycle if and only if the imaginary part of the action is increasing, 
	\begin{equation} \label{eq:TisIcondition}
		\mathcal{T}_{\tilde{h}} = \mathcal{I}_{\tilde{h}} \quad \Leftrightarrow  \quad a_{\tilde{h}} \le a_{\tilde{h}+1}  \,.
	\end{equation}
	To understand this we can consider the example in Figure~\ref{fig:onevar}, see also Figure~\ref{fig:ah}. When $\theta = \pi/2$, $a_{\tilde{h}} =- y_{\tilde{h}}^2/2$ is decreasing, and no thimble is an elementary cycle; the same is true when $\theta$ increases a bit to $0.55\pi$.  Now, when $\theta$ further increases to $0.7 \pi$, we see that the thimble with $\tilde{h} = 0$ (in black in Figure~\ref{fig:onevar}) has gone through a Stokes phenomenon. At the transition (somewhere between $\theta = 0.55\pi$ and $0.7\pi$), it must go to the blue saddle point with $\Tilde{h} = 1$. So $a_{0} - a_1 = 0$ at the transition by \eqref{eq:ImSeff}, and thus $a_{0} - a_1$ changes sign from $> 0$ (when $\theta = 0.55\pi$) to $< 0$ (when $\theta = 0.7\pi$). At the latter point, $\mathcal{T}_0$ becomes an elementary cycle. The same happens to  $\mathcal{T}_1$ between $\theta = 0.7\pi$ and $\theta = 0.85 \pi$, and so on. Finally, when $\theta = \pi$, $a_{\tilde{h}} = \pi \eta \tilde{h}$ is increasing with $\tilde{h}$, and all the thimbles are elementary cycles.
  
  Now, combining \eqref{eq:ImSeff-exp} and \eqref{eq:TisIcondition}, we find that  $\mathcal{T}_{\tilde{h}} = \mathcal{I}_{\tilde{h}} $ becomes valid when $\theta$ exceeds a threshold (which corresponds to the last Stokes wall for $\mathcal{T}_{\tilde{h}}$):
	\begin{equation}
		\mathcal{T}_{\tilde{h}} = \mathcal{I}_{\tilde{h}} 
		\quad \Leftrightarrow  \quad 
		\theta > \theta_{\tilde{h}} := \pi - \arctan\left( \frac{2 \pi \eta}{y_{\tilde{h}+1}^2 - y_{\tilde{h}}^2 } \right) \,. \label{eq:thetah}
	\end{equation}
	Now, recall that $y_{\tilde{h}}$ is the solution to $\tan(y/2) = -\eta y$ in the interval $(2\pi \tilde{h} - \pi,2\pi \tilde{h} + \pi )$, so $y_{\tilde{h}} \sim 2\pi \tilde{h} $ as $\Tilde{h} \to +\infty$. Plugging this into \eqref{eq:thetah} we may show that $\theta = \pi$ is an accumulation point of Stokes walls:
	\begin{equation}
		\pi -  \theta_{\tilde{h}}  \sim 
		\frac{ \eta}{2 \tilde{h} } \to 0 \,,\, \tilde{h} \to +\infty \,.
	\end{equation}
	In other words, an infinite number of Stokes phenomena  take place as we approach the time-like point. Hence, the thimble decomposition \eqref{eq:decom-thimble} of the Coulomb gas contour $(\im - \epsilon)\R$ is only true as a limit $\theta \to \pi$. In fact, for any $\theta < \pi$, $(\im - \epsilon)\R$ is a sum of a finite number  of thimbles, and that number diverges as $\theta \to \pi$. This accumulation of Stokes walls is the main outcome of our analysis. It is a feature of Coulomb gas effective action that is not present in the familiar Gamma function integral. In the latter, there is a unique Stokes wall at $\Re(n) = 0$, see the last paragraph of Section~\ref{sec:onesite}.

	The continuation we considered involve other Stokes phenomena as well, which we briefly comment on now. By symmetry, $\theta = 0$ is also an accumulation of the Stokes walls, however that does not affect the decomposition of the contour $\R$. Finally, $\theta = \pi/2$ is also a Stokes wall for all the thimbles $\mathcal{T}_{\tilde{h}}^{h_0}$, $\tilde{h} \ne 0 $. In fact, for a fixed $\Tilde{h}$, the Stokes phenomena of $\mathcal{T}_{\tilde{h}}^{h_0}$, viewed in the cover space, are essentially identical to those of the Gamma function integral thimbles described in Section~\ref{sec:onesite}. (Going to a zero of $\cosh$ is analogous to $\Re(\varphi)\to -\infty$; the integral converges there only on one side of the Stokes wall.)

 \begin{figure}
	     \centering
      \subfigure[]{
	     \includegraphics[scale=0.6]{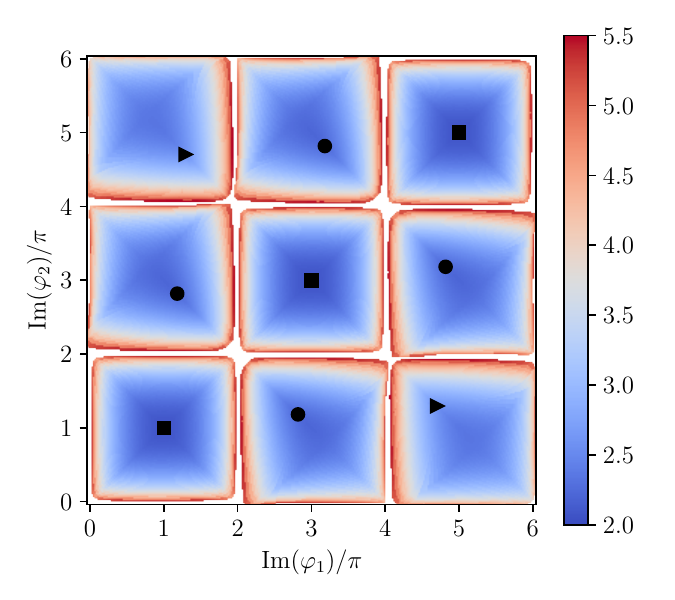}}
      \subfigure[]{\includegraphics[scale=0.6]{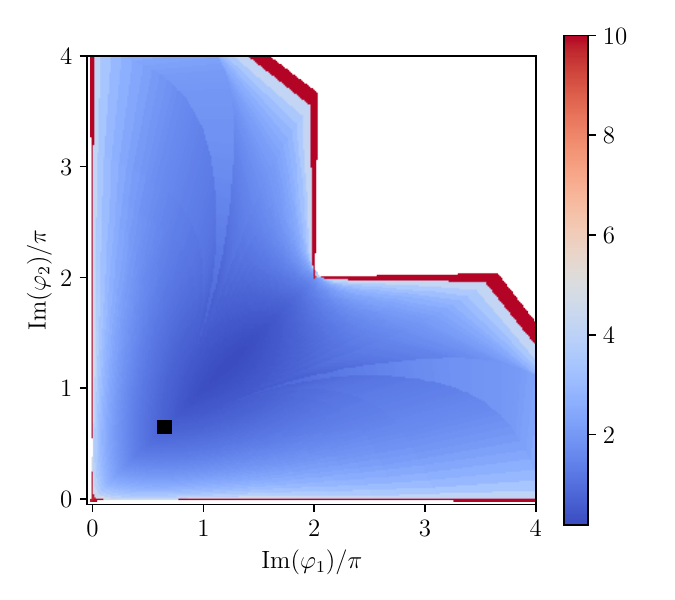}}
	     \caption{(a) Lefschetz thimbles of the two site action, with $\eta = 8$, and $b^2 = -1$, projected to the plane of $(\Im (\varphi_1) , \Im (\varphi_2))$. The solid markers indicate projection of the saddle points; the ones with the same type of marker are related by a shift of the zero mode $\varphi_0$. The color code indicate $\max_{i=1,2} \Re(\varphi_i)$. We observe that the thimbles are similar to the elementary cycles in that, the imaginary parts of the field are confined in boxes $\varphi_i \in [2\pi h_i, 2\pi (h_i+1)]$, $h_i \in \Z$. (b) A Lefschetz thimble with  $\eta=1$ and $b^2  = e^{-\im\theta}, \theta = 0.65 \pi $,  with the same color code above. It is no longer confined in a square box of side $2\pi$. This indicates that a Stokes phenomenon has taken place in the interval $\theta \in (0.65\pi, \pi)$. }
	     \label{fig:thimbles}
	 \end{figure}
	So far we used the effective action as a simpler proxy to the two-site action.  The saddle points of the latter are in one-to-one correspondence with those of $S_{\text{eff}}$. Indeed, using the expression of the action~\eqref{eq:S_phi0phid} in the coordinate system $(\varphi_0, \phid)$~\eqref{eq:twositezeromode-def}, it is not hard to check that if $(\varphi_1, \varphi_2)$ is a saddle point of $S$, $\phid = \varphi_1 - \varphi_2 = \varphi_{\tilde{h}}$ is a saddle point of $S_{\text{eff}}$, and  $\varphi_0 =  (\varphi_1 + \varphi_2) / 2$ satisfies $$ 
     \exp(\varphi_0) = \frac{\eta}{b^2} \, \mathrm{sech}\left(\frac{\phid_{\tilde{h}}}2 \right) \,.
     $$
    There are infinitely many such $\varphi_0$'s, which are specified by their imaginary part. The latter is of the form $ 2\pi h_0 + \theta $, such that $h_{1,2} = h_0 \pm \tilde{h}/2 \in \Z$. Hence, the saddle points of $S$ are also indexed by $(h_1, h_2) \in \Z^2$. In fact, it is easy to see from \eqref{eq:phid-trans-eq} that in the limit of $\eta \to \infty$, $\varphi_{\tilde{h}} \to 2\pi \im \tilde{h}$, so the saddle points form literally a square lattice $(\varphi_1 , \varphi_2) \to (2\pi h_1+ \theta, 2\pi h_2+ \theta)$, see Fig~\ref{fig:thimbles}-(a). Now, to each saddle point is associated a thimble $\mathcal{T}_{h_1,h_2}$. By definition, it can be generated by the solutions to the upward flow equation
    \begin{equation} \label{eq:upflow-two-site}
        \overline{\partial_s \varphi_i(s)} =  \partial_{\varphi_j} S(\varphi_1, \varphi_2) \,,\, i = 1, 2 \,,
    \end{equation}
    that tend to the saddle point as $s\to -\infty$. Note that $\mathcal{T}_{h_1,h_2}$ is a manifold of real dimensional $2$ embedded in $\C^2$, which has real dimension $4$. This makes the thimbles of the two-site action relatively difficult to visualize and understand analytically. Thus, we resort to a numerical study, see Appendix~\ref{sec:num} for methods. We construct the thimbles $\mathcal{T}_{h_1,h_2}$ (with small $\tilde{h} = h_1 - h_2$) explicitly by integrating the upward flows. 
    
\begin{figure}
        \centering
        \includegraphics[scale=0.9]{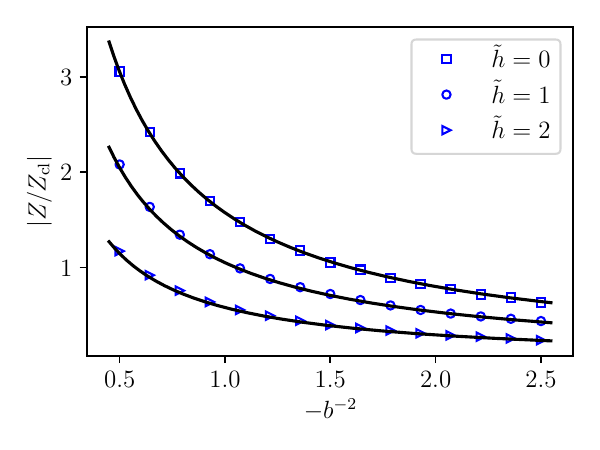}
        \caption{Comparing the two-site path integral evaluated on Lefschetz thimbles (markers) \textit{vs.} that evaluated on elementary cycles (continous curves), for various values of $b^2 < 0$ and $\tilde{h}$; we chose $\eta = 3$. The former is calculated \textit{ab initio}, by constructing the thimble numericaly, see Appendix~\ref{sec:num} for methods. The latter is calculated by using the effective action \eqref{eq:ZandZCG}, and numerically evaluating the integral $\widetilde{Z}(\mathcal{I}_{\tilde{h}})$. We divide the results by $Z_{\text{cl}}$, which is the value of the action on the saddle point. The agreement supports the claim that the Lefschetz thimbles are equivalent to the elementary cycles when $b^2 < 0$. }
        \label{fig:thimble_integral}
    \end{figure}

    As a result, we find that thimbles are equivalent to the elementary cycles when $b^2 < 0$ and $\eta > 0$. This can be seen qualitatively by plotting their projection in to the plane $(\Im (\varphi_1), \Im (\varphi_2))$, see Figure~\ref{fig:thimbles} (a). The projection of $\mathcal{T}_{h_1,h_2}$ appears to be contained in the box $[2\pi h_1, 2\pi (h_1+1) ] \times [2\pi h_2, 2\pi (h_2+1) ]$, just as that of the elementary cycle $\mathcal{U}_{h_1, h_2}$. 
    As a quantitative confirmation, we also evaluated the path integral on the thimbles numerically,
    $$ Z(\mathcal{T}_{h_1,h_2}) = \int_{\mathcal{T}_{h_1,h_2}}  e^{-S(\varphi_1, \varphi_2)} \mathrm{d} \varphi_1 \mathrm{d} \varphi_2 \,, $$
    and compared the result to that on the corresponding elementary cycle, calculated using the integral representation~\eqref{eq:ZU2-int}. We find a good agreement, indicating that
   \begin{equation}
       Z(\mathcal{T}_{h_1, h_2}) =  Z(\mathcal{U}_{h_1, h_2})  \,,\, \eta > 0 \,,\, b^2 < 0 \,.
   \end{equation}
   See Figure~\ref{fig:thimble_integral} for a plot. As a consequence, we can conclude that the continuation of the $c\ge 25$ path integral to $c\le 1$ is an infinite sum of thimbles:
   \begin{equation} 
       Z(\R^2) = \sum_{h_1\ge 0 , h_2\ge 0}  Z(\mathcal{T}_{h_1, h_2}) \,,\quad  \widehat{Z} = 
       \sum_{\min(h_1,h_2) =  0}  Z(\mathcal{T}_{h_1, h_2}) \,,
   \end{equation}
   for $b^2 < 0, \eta > 0$. In contrast, when $b^2 > 0, \eta > 0$, $Z(\R^2)$ is the integral over a single thimble. So, in these cases, the thimbles of the two-site action have the same behavior as those of the (one-variable) effective action. We also observe that the thimbles of the two-site action undergo Stokes phenomena inside the interval $\theta \in (\pi/2,\pi)$, where they become no longer confined in boxes, see Figure~\ref{fig:thimbles} (b). We checked that the Stokes wall positions $\theta_c$ are consistent with the prediction \eqref{eq:TisIcondition} obtained from the effective-action analysis. In summary, our numerical study of the two-site thimbles indicates that the Stokes phenomena of the effective action represent faithfully those of the two-site action.


	\section{Lattice Liouville path integral}\label{sec:gen}
	The Liouville action on a connected compact 2D surface $\Sigma$ without boundary is defined as~\footnote{Often, one parametrizes the action a bit differently, with an interaction term $e^{b\phi}$ and a $b$-independent kinetic term. This parametrization is related to the ours by a change of variable $\varphi = b \phi$.}
	\begin{equation}\label{eq:S-liouville}
		S =  \int_{\Sigma} \left( \frac{1}{16\pi b^2} (\nabla \varphi)^2 + \frac{1}{8\pi} (1 + b^{-2}) R \varphi + \mu e^{\varphi} \right) \mathrm{d} \mathcal{A} \,.
	\end{equation}
	Here, $\mathrm{d} \mathcal{A}$ is the area element of the surface, $R$ its scalar curvature, and $b$ is related to the central charge $c$ by 
	\begin{equation} \label{eq:cbrelation-real}
		c = 1 + 6 \left(b + b^{-1} \right)^2 \,.
	\end{equation}
	$\mu$ is known as the cosmological constant, and affects correlation functions in a simple way. The standard observables are correlation functions between ``vertex operators'', which are exponential of the field. So the general form of a Liouville path integral is the following: 
	\begin{equation}
		\int [\mathcal{D}\varphi] \, e^{-S[\varphi] + a_1 \varphi(z_1)  + \dots +  a_n \varphi(z_n) }
	\end{equation}
 Here, the meaning of $\int [\mathcal{D}\varphi]$ is well-understood only when $b^2 > 0$, or $c\ge25$. We shall consider a naive regularization of this action on a 2D planar lattice.  The field is defined on the faces of the lattice: $$ \varphi(z) \to \varphi_r \,. $$
	The kinetic term will be replaced by a sum over lattice bonds:
	\begin{equation}
		\int_{\Sigma}\frac{1}{16\pi b^2} (\nabla \varphi)^2  \mathrm{d} \mathcal{A} \to
		\sum_{\left< r,r' \right>} \frac{K_{rr'}}{2b^2} (\varphi_r - \varphi_{r'})^2  \,.
	\end{equation}
	Here and below we denote a bond by the two faces $r,r'$ it separates. The coefficients $K_{rr'} > 0$ are to be chosen appropriately as a function of the lattice to reproduce the continuum kinetic term in the scaling limit. For example, on a square lattice we may choose $K_{rr'} = 1/(8\pi)$ (note that in 2D, the Gaussian free field has scaling dimension $0$, so $K$ does not diverge with the lattice spacing in the continuum limit) The interaction term $\propto \mu $ becomes 
	\begin{equation}
	\mu 	\int_{\Sigma}   e^{\varphi}  \mathrm{d} \mathcal{A} \to \sum_r \mu_r e^{\varphi_r} \,,
	\end{equation}
	where $\mu_r = \mu A_r$ and $A_r$ is the area of the face $r$. Finally,  both the coupling to curvature and the vertex operator insertion can be accounted for by a position-dependent source $\eta_r$:
	\begin{equation} \label{eq:aQandeta}
		- \sum_{i} a_i \varphi(z_i)  + \int_{\Sigma}   \frac{1}{8\pi} (1 + b^{-2}) R \varphi \mathrm{d} \mathcal{A}  \to   - \sum_{r} \frac{\eta_r}{b^2} \varphi_r \,.
	\end{equation}
	In terms of $\eta_r$, the Seiberg bound at $c\ge 25$ is again given by
	\begin{equation}
		\sum_{r} \eta_r > 0 \,.\label{eq:seiberg-gen}
	\end{equation}
	In summary, we have a lattice Liouville action:
	\begin{equation}
		S[\varphi_r] =   \sum_{\left< r,r' \right>} \frac{K_{rr'}}{2b^2} (\varphi_r - \varphi_{r'})^2 + \sum_{r} \left[ \mu_r e^{\varphi_r} - \frac{\eta_r}{b^2} \varphi_r \right] \,. \label{eq:lattice-action}
	\end{equation}
	When the Seiberg bound is satisfied, the $c\ge25$ Liouville theory can be defined by the continuum limit of a lattice path integral where every $\varphi_r$ is integrated over the real line:
	\begin{equation}
		Z(\R^N) = \int_{\R^N} \prod_r  \mathrm{d} \varphi_r	e^{-S[\varphi_r] } \,, \label{eq:latticeZ-c25}
	\end{equation} 
	where $N$ is the number of faces of the planar lattice. Based on rigorous results~\cite{david16} and numerical tests~\cite{lft}, we expect the lattice path integral has a continuum limit described by the $c\ge25$ Liouville theory provided the Seiberg bounds~\cite{seiberg90} and also $b < 1$. Our goal here is to calculate the analytical continuation of the lattice path integral as $b^2$ rotates clockwise from $\R_+$ to $\R_-$, that is, we let 
	$$ b^{2} = |b^2| e^{-\im \theta} \,,\, \theta: 0 \to \pi \,, $$ 
    while the $\eta_r$'s are kept fixed. In the continuum theory, this amounts to fixing the surface geometry (hence $R$), and rotating $\eta_r$ along with $b^2$ so that $\eta_r / b^2$ is kept fixed; in particular, $\eta_r$ will change sign as does $b^2$. 
	
	\subsection{The zero mode approach}
	It is helpful to review first the zero mode approach, which is well-understood when  $c\ge25$, and appreciate the issues involved in its analytical continuation. Like in the two-site model, we write the Liouville field as a sum of a zero mode $\varphi_0$ and a fluctuating field $\phid_r$ that has vanishing sum:
	\begin{equation}
		\varphi_r = \varphi_0 + \phid_r \,,\, \sum_r \phid_r = 0 \,.
	\end{equation}
	To evaluate the $c\ge25$ path integral, we can integrate over the zero mode: for any fixed $\phid$, the zero-mode integral one-site Gamma function integral \eqref{eq:gamma-integral-line}. Introducing the notations\footnote{$A[\phid]$ is sometimes called $Z[\phid]$ since when $b > 0$ it can be viewed as the partition of a particle in a random potential $\propto - \phid_r$. We shall avoid this notation here as $Z$ stands for the Liouville path integral. }
	\begin{align}
		&n := - \sum_r \frac{\eta_r}{b^2} \,,\,  \label{eq:defn} \\
		&A[\phid] := \sum_r \mu_r e^{\phid_r} \,, \label{eq:defA}
	\end{align}
	we have 
	\begin{align}
		&\int_{\R} \mathrm{d} \varphi_0 \; e^{-S[\varphi]} \nonumber \\
		=& 
		\exp \left( -
		\sum_{\left< r,r' \right>} \frac{K_{rr'}}{2b^2} (\phid_r - \phid_{r'})^2 + \sum_r \frac{\eta_r}{b^2} \phid_r \right)  \int_{\R} \mathrm{d} \varphi_0 \exp\left( - n \varphi_0 - e^{\varphi_0} A[\phid] \right)  \nonumber \\
		=&  \Gamma\left( - n \right) \exp(-S_{\text{eff}} [\phid] ) \,, \label{eq:zeromode-gen}
	\end{align}
	where the resulting effective action is:
	\begin{align}
		S_{\text{eff}}[\phid] =  \frac{K_{rr'}}{2b^2} (\phid_r - \phid_{r'})^2 - \sum_r \frac{\eta_r}{b^2}\phid_r - n \ln A[\phid]  \,. \label{eq:Seff-gen}
	\end{align}
	Thus the $c\ge25$ Liouville theory becomes a path integral over the fluctuating modes 
	\begin{equation} \label{eq:ZRnfluct}
		Z(\R^N) =  \Gamma(-n)  \int [\mathcal{D} \phid] \exp(-S_{\text{eff}}[\phid])\,,\, 
		\int \mathcal{D} \phid = \int_{\R^N} \prod_r \mathrm{d} \phid_r \delta\left(\sum_r \phid_r \right)  \,.
	\end{equation}
   We see that when $n = 0,1,2,3, \dots$, the Gamma function has a pole. We can remove it by defining
   \begin{equation}
   	  \widehat{Z} := 	Z(\R^N) (1 - e^{- 2\pi \im n  }) = \frac{2\pi \im e^{-\im \pi n }}{\Gamma(1+n)}  \int [\mathcal{D} \phid] \exp(-S_{\text{eff}}[\phid])\,, \label{eq:widehatZ-gen1}
   \end{equation}
  similarly as in the two-site toy model above. We can also think of $\widehat{Z}$ as being obtained by integrating over the zero mode over an elementary cycle $\mathcal{U}_{0}$. 
   
    It remains to consider the continuation of the fluctuation mode integral in the right hand side of \eqref{eq:widehatZ-gen1}. We may attempt to think of it as evaluating an average over a Gaussian free field (defined by the first two terms of \eqref{eq:Seff-gen}), yet of a complex observable $ A[\phid]^n$. However, as $b^2$ rotates in the complex plane to the region where $\Re(b^2) < 0$, we will have to rotate of the $\phid$ integral away from the real hyperplane, into a submanifold of $\C^N$. There, $ A[\phid]^n $ is multi-valued and we need to specify the integration cycle with respect to the branch cuts. We have encountered a particularly mild version of this problem in the two-site toy model above. In that model, $\phid$ reduces to a single variable, $A[\phid] = 2 \cosh(\phid / 2)$. Its zeros are purely imaginary, so that it is possible to deform the contour to $\phid \in (\im - \epsilon) \R$. In general, it is a formidable task to determine where $A[\phid] = 0$, and it does not imply necessarily $\phid_r \in \im \R$. So we cannot deform the fluctuating mode integral in \eqref{eq:ZRnfluct} to an infinitesimally tilted deformation of $(\im \R)^N$.  
	
	One exception is when $n \in \mathbb{N}$ is a nonnegative integer, in which case $A[\phid]^n$ is single-valued and we can integrate over $\phid_r \in \im \R $ when $b^2 < 0$. It is then more convenient to write $b = \im \beta$, $\phid_r = \im \beta \psi_r $, $\beta \in \R$, in terms of which we have
	\begin{equation}  \label{eq:CGgen0}
	 \widehat{Z} / c = \frac{1}{n!} 
		\int_{\R^N} \left[\mathcal{D}  \psi \right] \exp(- S_0[\psi] ) e^{-\im  \sum_r {\eta_r \psi_r}/{\beta}} A[\im \beta \psi]^n \,,\, b^2 <  0\,, 
	\end{equation}
	where $c= 2\pi \beta^N \im^{N+1} (-1)^n$ is an unimportant constant, and $S_0$ is a non-interacting action:
	\begin{equation}
		S_0[\psi] = \frac{K_{rr'}}{2} (\psi_r - \psi_{r'})^2  \,.
	\end{equation}
   Hence, we can view $\psi$ as a lattice Gaussian free field defined by the action $S_0$, and denote its Green function as $G_{rr'} = \left< \psi_r \psi_{r'} \right>_{S_0}$. Then, expanding the ``observable'' $A[\im \beta \psi]^n $ in \eqref{eq:CGgen0} and applying the Wick theorem, we obtain the (lattice) Coulomb gas representation:
   \begin{equation}  \label{eq:ZhatCGgen}
   	     \widehat{Z}/c  = Z_{\text{GFF}}  \sum_{\sum_r n_r = n}    \frac{1}{n_r!} \exp\left[  -\frac12 \sum_{r r'} \left( n_r \beta - {\eta_r}/{\beta} \right) G_{rr'}   \left( n_{r'} \beta - {\eta_r}/{\beta} \right)  \right]
   \end{equation}
where the sum is over configurations with $n$  ``screening charges'' of the same type as that appearing in the action~\eqref{eq:S-liouville}, and 
$$Z_{\text{GFF}} :=  \int [\mathcal{D} \psi] \exp(- S_0[\psi] )  \, $$
is the partition function of the Gaussian free field. In the continuum limit, \eqref{eq:ZhatCGgen} becomes  an integral over $n$ positions on the surface $\Sigma$. 

In summary, the $c\ge25 \to c\le1$ continuation of the Liouville path integral gives rise to the Coulomb gas integral at special values of the parameters, which are known as satisfying  a ``charge neutrality'' condition. Yet, for generic parameters, the zero-mode approach does not seem to lead anything tractable.

	\subsection{Elementary cycle decomposition}\label{sec:elementary-gen}
	In this section, we apply the methods of Section~\ref{sec:elementary-twosite} to the general lattice Liouville path integral. We shall assume that the charges of the vertex operators are real, $a_i \in \R$, and that the Seiberg bound is satisfied. Equivalently, in terms of $\eta_r$, we assume: 
	\begin{equation}
		\eta_r \in \R \,,\, \sum_{r} \eta_r > 0 \,.
	\end{equation}
	In particular, some $\eta_r$ can be negative. 
	
   	We first decouple the kinetic term in \eqref{eq:lattice-action} by introducing one variable $\chi_{rr'}$ for each edge $rr'$ (note that we denote an edge by the two faces it separates). We shall view $\chi$ as a discretized one form and set $
	\chi_{rr'} \equiv -\chi_{r'r} $ by convention. As a result, we have  
	\begin{align}
		\exp({-S[\varphi]}) &= \int \prod_{\left< rr'\right>} \frac{e^{\frac{b^2 \chi_{rr'}^2}{2K_{rr'}}} \mathrm{d}\chi_{rr'}}{\im \sqrt{2\pi K_{rr'} / b^2}} \prod_r \exp( - n_r \varphi_r  - \mu_r e^{\varphi_r} ) \label{eq:decoupled-action-gen} \\
		n_r &= -\frac{\eta_r}{b^2} + (\mathbf{d} \chi)_r \,.   \label{eq:nrdef}
	\end{align}
	Here, $ (\mathbf{d}  \chi)_r$ is the discrete exterior derivative (curl) of the one form $\chi$: $ (\mathbf{d}  \chi)_r = \sum_{\left< rr' \right>} \chi_{rr'} $. The contour for $\chi_{rr'}$ can be arbitrarily chosen as long as the Gaussian integral is convergent. Equations~\eqref{eq:decoupled-action-gen} and \eqref{eq:nrdef} are direct generalization of \eqref{eq:decouple2} and \eqref{eq:n1n2} of the two-site toy model. The $\varphi$ integral is again decoupled into a product of one-site Gamma function integrals, upon a simple change of variable $\varphi'_r = \varphi_r + \ln \mu_r$ (note that $\mu_r$ are positive constants and we choose the branch of the log with $\ln \mu_r \in \R$).

	We now use \eqref{eq:decoupled-action-gen} and \eqref{eq:nrdef} to derive an integral representation of $Z(\R^N)$ \eqref{eq:latticeZ-c25} for $b^2 > 0$, analogous of \eqref{eq:ZR2-int} for the two-site model. Here, we need to be a bit more careful in choosing the contour of $\chi_{rr'}$. When $b^2 > 0$, the contour can be vertical. Yet, $\chi_{rr'} \in \im \R$ for all edges $rr'$ is not always a good choice: since $\eta_r$ is not always positive, $-\Re(n_r) = \eta_r / b^2$ can be negative, which leads to a divergent integral $$ 
	\int_{\R}  \exp( - n_r \varphi_r  - \mu_r e^{\varphi_r} ) \mathrm{d} \varphi_r 
	$$ as we integrate over $\varphi_r$'s in  \eqref{eq:decoupled-action-gen}. However, thanks to the Seiberg bound $\sum_r \eta_r > 0$, we can shift the contour $\chi_{rr'}$ horizontally, 
	\begin{equation}
		\chi_{rr'} \in \frac{\sigma_{rr'}}{b^2} + \im \R \label{eq:chicontour-c25-gen} \,,\, b^2 > 0\,.
	\end{equation}
	where $\sigma_{rr'} \in \R$ are chosen such that 
	\begin{equation}\label{eq:sigmacondition}
		\mathrm{Re}(n_r) = \frac{(\mathbf{d} \sigma)_{r} - \eta_r }{b^2}   <  0 \,,\, b^2 > 0 \,,
	\end{equation}
	for all $r$. (To do this, we can let $\eta'_r = \eta_r - \frac1N \sum_{r} \eta_r$ where $N$ the number of faces, so that $\sum_r \eta'_r = 0$. Then there exists $\sigma$ so that $\mathbf{d} \sigma = \eta'_r $, and thus $\mathrm{Re}(n_r) = - \frac1{N b^2} \sum_{r} \eta_r < 0$.) 
	Then, we can integrate over $\varphi_r$ and obtain the following:
	\begin{equation}\label{eq:ZRn-int-gen}
		Z(\R^N) = \int \prod_{\left< rr'\right>} \frac{e^{\frac{b^2 \chi_{rr'}^2}{2K_{rr'}}} \mathrm{d}\chi_{rr'}}{\im \sqrt{2\pi K_{rr'} / b^2}} \prod_{r}  \mu_r^{n_r} \Gamma(-n_r)  \,.
	\end{equation}
	We can use this formula to continue $Z$ as we change $b^2$, by deforming smoothly the $\chi_{rr'}$ contours \eqref{eq:chicontour-c25-gen} in a way such that $-n_r$ avoids the poles of the Gamma function for all $r$. One such way is to let 
	\begin{equation} \label{eq:chirr-contour-gen}
		\chi_{rr'} \in {\sigma_{rr'}}/{b^2} + e^{\im \omega} \R \,,\, 
	\end{equation}
	and change $\omega$ smoothly along with $\theta$ so that the Gaussian integral of $\chi_{rr'}$ converges and that $\omega \in [\pi/2, \pi]$. The latter condition ensures that 
	$$ n_r \in ((\mathbf{d} \sigma)_{r} - \eta_r)/b^2  + e^{\im \omega} \R  $$
	belongs to a line that never crosses $[0, \infty)$, and thus stays away from all poles of $\Gamma(-n_r)$ for any $\theta \in (0, \pi)$ (Figure~\ref{fig:ni-contour} still applies, provided we replace $\eta_r$ by $\eta_r - (\mathbf{d} \sigma)_r$). We conclude that $Z(\R^N)$ can be continued up to the vincinity of the $c\le 1$ line. 
	
	The notion of elementary cycles also generalizes straightforwardly to a general lattice. Here an elementary cycle is indexed by a ``height function'' $\mathbf{h}$ that assigns an integer $h_r$ to every lattice face $r$, and defined again as a product: 
	\begin{equation} \label{eq:Udef-gen}
		\mathcal{U}_{\mathbf{h}} = \prod_r \mathcal{U}_{h_r}(\varphi_r) 
	\end{equation}
	where the single-variable elementary cycle $\mathcal{U}_h$ is defined in \eqref{eq:Undef}. The definition \eqref{eq:Udef-gen} is identical to the two-site one~\eqref{eq:Udef-twosite}. by the same argument below \eqref{eq:Udef-twosite}, the path integral converges on any elementary cycle, and admits an integral representation similar to \eqref{eq:ZU2-int}:
	\begin{equation} \label{eq:ZU-int-gen}
		Z(\mathcal{U}_{\mathbf{h}}) := \int_{\mathcal{U}_{\mathbf{h}}} [\mathcal{D} \varphi] e^{-S[\varphi]} = 
		\int \prod_{\left< rr'\right>} \frac{e^{\frac{b^2 \chi_{rr'}^2}{2K_{rr'}}} \mathrm{d}\chi_{rr'}}{\im \sqrt{2\pi K_{rr'} / b^2}} \prod_{r}  \frac{2 \pi \im \mu_r^{n_r} e^{-(2h_r+1) n_r \pi \im}}{\Gamma(1 + n_r)} \,.
	\end{equation}
	In particular, $Z(\mathcal{U}_{\mathbf{h}})$ depends on the global shift of $\mathbf{h} \to \mathbf{h} + h_0$ ($h_0 \in \Z$) in a simple way 
	\begin{equation}\label{eq:ZUshift-gen}
		Z(\mathcal{U}_{\mathbf{h} + h_0}) =
		Z(\mathcal{U}_{\mathbf{h}}) \exp\left( 2 \pi\im h_0 \sum_{r} \eta_r / b^2 \right)  = Z(\mathcal{U}_{\mathbf{h}}) \exp\left( - 2 \pi\im n h_0\right)  
	\end{equation}
 where we recall $n =  -\sum_r \eta_r / b^2$~\eqref{eq:defn}. To see \eqref{eq:ZUshift-gen} from \eqref{eq:ZU-int-gen}, note that $\sum_r n_r = \sum_r ( (\mathbf{d} \chi)_r - \eta_r / b^2) = -\sum_r \eta_r / b^2$ since the curl has a vanishing integral. 
	
	We are now ready to state the main result of this section. The analytically continued $c\ge25$ path integral is the sum of all elementary cycles indexed by a non-negative height function at a vicinity of the $c\le 1$ line:
	\begin{equation}\label{eq:ZRn-decomp}
		Z(\R^N) = \sum_{\mathbf{h}: h_r \ge 0}  Z(\mathcal{U}_{\mathbf{h}}) \,,\, \pi / 2 < \theta  < \pi.
	\end{equation}
	The proof is similar to the two-site case. Indeed, for $\theta > \pi/2$, the integral formula~\eqref{eq:ZRn-int-gen} for $Z(\R^N)$ can have a contour \eqref{eq:chirr-contour-gen} with $\omega = \pi$, that is, a horizontal line $\chi_{rr'} \in \sigma_{rr'} / b^2 + \R $ so that for any $r$, we have always
	$$ \Im(n_r) = \Im \left[\frac{(\mathbf{d} \sigma)_r - \eta_r}{b^2} \right] =  |b|^{-2} \sin (\theta)((\mathbf{d} \sigma)_r - \eta_r) < 0 \,,\, \theta \in (0, \pi) \,,  $$
	where we recall $b^2 = |b^2| e^{-\im \theta}$ and used \eqref{eq:sigmacondition} (see Figure~\ref{fig:ni-contour} with $\eta_i$ replaced by $\eta_r - (\mathbf{d} \sigma)_r$). Therefore, by \eqref{eq:gamma-decomp}, for any $(\chi_{rr'})$, every integral over $\varphi_r \in \R$ \eqref{eq:ZRn-int-gen} is a convergent sum over all elementary cycles $ \mathcal{U}_{h_r} $ with $h_r \ge 0$. This proves \eqref{eq:ZRn-decomp}. 
	Using \eqref{eq:ZRn-decomp} and \eqref{eq:ZUshift-gen} it is not hard to check that that the modified path integral $\widehat{Z}$ defined by \eqref{eq:widehatZ-gen1} is a sum over elementary cycles with vanishing minimal ``height'':
	\begin{equation}\label{eq:Zhat-gen}
		\widehat{Z} = \sum_{\min h_r = 0}  Z(\mathcal{U}_{\mathbf{h}})  \,.
	\end{equation}
   Equations \eqref{eq:ZRn-decomp} and \eqref{eq:Zhat-gen} are the main outcome of our analysis. In the rest of this section,  we briefly discuss two interpretations of these results, leaving a more thorough study to future work.
\begin{figure}
    \centering
    \includegraphics[scale=0.75]{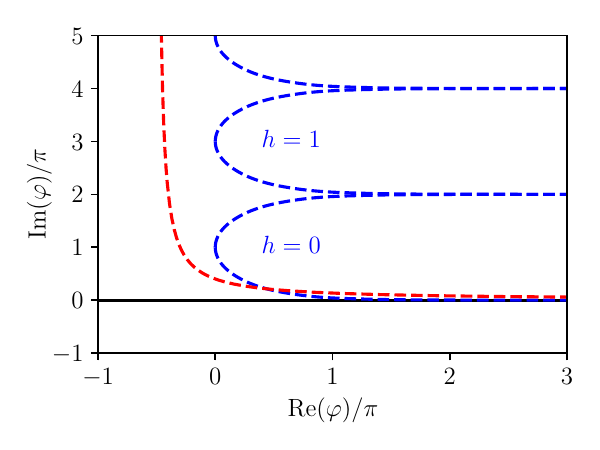}
    \caption{The L-shape contour (red dashed curve) is a sum of elementary cycles (blue dashed curves) with $h > 0$, and is obtained as a result of continuing the $c\ge25$ Liouville path integral on $\R$ (black solid curve) to the vicinity of $c\le 1$.  }
    \label{fig:l-contour}
\end{figure}

One equivalent way of \eqref{eq:ZRn-decomp} it to say that, to continue the $c\ge25$ path integral to the vicinity of $c \le 1$, it suffices to deform the contour of all field components from $\varphi_r \in \R$ to $\varphi_r \in \mathcal{L}$, where $\mathcal{L}$ is a L-shaped contour in $\C$ that connects $+\im \infty$ to $+\infty$, see Figure~\ref{fig:l-contour}. Such a contour will have implications for the analytical continuation of Liouville quantum mechanics~\cite{mcelgin,fredenhagen-shomerus,Brody_2014,Kapec:2020xaj}. Liouville quantum mechanics for $b^2< 0$ has been considered as an example of a non-Hermitian, PT symmetric~\cite{bender} quantum theory. However, the Hilbert space corresponding to the integration contour we obtained has not been considered. 

The discrete sum over the height function in \eqref{eq:ZRn-decomp} and \eqref{eq:Zhat-gen} suggests a statistical interpretation in terms of loop models. An $O(n)$ lattice loop model can be defined by a partition sum over non-intersecting loop ensembles on a planar lattice, with a fugacity $n$ coupled to the number of loops. (The fugacity coupled to the total loop length is adjusted so that the system is critical.) It is common practice to orient the loops, and assign a complex weight $e^{\pm \im \pi e_0}$ to each of the orientations, such that $n  = 2\cos(\pi e_0)$. Then, the oriented loop configuration defines a discrete height field $h_{\mathbf{r}}$ on the faces, \textit{up to an overall shift (zero mode)}, such that the oriented loops are the isotopic lines of $h_{\mathbf{r}}$: whenever one crosses a loop from its left to its right, $h_{\mathbf{r}} $ increments by $1$. Well-known arguments~\cite{nienhuis,kondev,estienne2015correlation} show that this height field is described by the $c\le1$ Liouville action such that $c = 1 - 6 (b-1/b)^2$ and 
$$ e_0 =  1 - b^{-2}   \,,\, b \in [1/\sqrt{2}, \sqrt{2}] \,. $$ 
In particular, each $n \in [-2,2)$ correspond to two values of $c$, which describe the dilute and dense fixed points, respectively. This identification allowed to obtain critical exponents by simple free field calculations. These calculations are feasible partly because the correlation functions in question depend only on the difference between $h_{\mathbf{r}}$ evaluated at two points, and one does not need to specify the zero mode. This is not the case for generic correlation functions beyond two points: in addition to summing over the loop configurations, we need to specify the zero mode so that the vertex operators $e^{\alpha \varphi_{\mathbf{r}}} \to e^{\alpha 2\pi \im h_{\mathbf{r}}}$ can be evaluated without ambiguity. For instance, the loop model interpretation of the $c\le 1$ Liouville structure constant put forward in Ref.~\cite{IJS} involves a nontrivial prescription of the zero mode. If the three-point function is evaluated  on $z_1, z_2, z_3$, then $h_{\mathbf{r}} = 0$ for $\mathbf{r} \in G$, where $G$ is the unique region such that such that for any $i \in \{1, 2, 3\}$, $G$ is adjacent to a loop that surrounds $z_i$. It is unclear how to generalize this geometric prescription to four point functions.

Now, our result~\eqref{eq:ZRn-decomp} provides an interesting new prescription if we identify the integer-valued height function of the loop model with that indexing the elementary cycles. We should sum over all zero modes such that $\min_{\mathbf{r}} h_{\mathbf{r}} \ge 0$ in order to calculate the continuation of the $c\ge 25$ path integral, or fix the zero mode such that $\min_{\mathbf{r}} h_{\mathbf{r}} = 0$ to calculate the reduced continuation $\widehat{Z}$. 
We note that our prescription is different from that of \cite{IJS} on the lattice, and we expect this difference to persist in the continuum limit. Indeed, our prescription is intended to reproduce the analytical continuation of the $c\ge25$ Liouville correlation functions, while the latter gives rise to $c\le 1$  Liouville correlation functions. 
Identifying the height functions is reasonable \textit{in the continuum limit}. On each individual elementary cycle, fluctuation of the field $\varphi$ is suppressed: $\Im (\varphi_r)$ is confined in an interval of length $2\pi$, and $\Re(\varphi_r) \to +\infty$ along the elementary cycle is prohibited energetically. As a result, critical fluctuations can only result from the sum over $h_{{r}}$'s, weighted by an elastic energy $\propto \sum_{rr'} (h_{{r}} - h_{{r}'})^2$ plus source terms $\propto \eta_r h_r$ (in other words, we approximate $\varphi_r \approx (2 h_r + 1) \pi \im$ in an elementary cycle). Such a partition sum is equivalent to that of the loop-model height functions, up to irrelevant perturbations, e.g., those restricting the height difference between neighboring faces~\cite{nienhuis,kondev}.



\section{Conclusion}
We considered the problem of analytically continuing the lattice Liouville path integral~\eqref{eq:latticeZ-c25} from $c\ge25$ to generic values of $c$, focusing on the vicinity of $c\in (-\infty, 1]$. We showed that the result~\eqref{eq:ZRn-decomp} can be written as an discrete sum over elementary cycles, defined in \eqref{eq:Udef-gen}. These elementary cycles are a straightforward generalization of the inverse Gamma function contours~\eqref{eq:Undef}, appear to be a useful basis of the integration cycles of the Liouville action. We illustrated the convenience of this basis, compared to the Lefshetz thimbles, with a detailed analysis of the two-site toy model (Section~\ref{sec:twosite}), which features involved Stokes phenomena. 

The progress reported in this work is incremental in nature, since we considered exclusively finite lattices, and have not begun to address the elephant in the room --- the continuum limit. However, what is known about the bootstrap solutions suggests the following naive conjecture. as long as $c \notin (-\infty, 1]$ the continuum limit of \eqref{eq:ZRn-decomp} is described by the DOZZ-bootstrap solution without discrete terms, as long as we supposed the Seiberg bounds at the starting point of the continuation, and perform the continuation by rotation $b^2$ as was done above. On the other hand, we expect that the singularities that appear in the $c\to (-\infty, 1]$ limit of the DOZZ structure constants should emerge in the continuum limit of our proposal. It will be interesting to observe this in a numerical implementation of our proposal. For this purpose, a promising strategy is to further develop the statistical interpretation suggested by the discrete sum \eqref{eq:ZRn-decomp}, which is different from the standard loop model observables. 
	
	\appendix
	
\section{Proof of \eqref{eq:ZandZCG}}\label{sec:proofZandZCG}

In this appendix we show the formula \eqref{eq:ZandZCG} that relates an elementary cycle of the two-site toy model to one of the effective action. By analyticity we can assume that $\Re(\eta / b^2) > 0$ so that $Z(\R^2)$ converges. Then, using \eqref{eq:Udef-twosite} and \eqref{eq:Undef}, we can write an elementary cycle of the two-site model as a linear combination of shifted $\R^2$'s:  
\begin{equation} \label{eq:A1}
    Z(\mathcal{U}_{h_1, h_2}) = Z(\R_{h_1, h_2}) - Z(\R_{h_1+1, h_2}) + Z(\R_{h_1+1, h_2+1}) - Z(\R_{h_1, h_2+1}) \,,
\end{equation}
where 
\begin{equation}
    \R_{h_1, h_2} := \{\varphi_1 \in \R + 2\pi \im h_1 \,,\,\varphi_2 \in \R + 2\pi \im h_2  \}\,.
\end{equation}
Switching to the coordinate system $\varphi_0 = (\varphi_1 + \varphi_2)/2, \phid = \varphi_1 - \varphi_2$, and $h_0 = (h_1 + h_2) / 2, \tilde{h} = h_1 - h_2$, we have 
\begin{equation}
     \R_{h_1, h_2} = \{ \varphi_0 \in  \R + 2\pi \im h_0 \,,\, \phid \in \R + 2\pi \im \tilde{h}  \}\,.
\end{equation}
In the four terms of \eqref{eq:A1}, $\Im(\phid) = \tilde{h}, \tilde{h} + 1 ,\tilde{h} , \tilde{h} -1  $, and $\Im (\varphi_0) = h_0, h_0 + \frac12, h_0 + 1, h_0 + \frac12$, respectively. As we integrate out the zero mode in each term (with $\phid$ fixed), we will get a Gamma function $\Gamma(2\eta/b^2)$ and a phase $e^{\im 2\eta/b^2 \times \Im \varphi_0}$. Yet, the latter can be can be absorbed by choosing the appropriate branch of the logarithm in the effective action~\eqref{eq:Seff}. We can write:
\begin{equation}\label{eq:ZU1U2-contour}
   Z(\mathcal{U}_{h_1, h_2}) = \Gamma(2\eta / b^2) \left( \int_{\R + 2\pi \im \tilde{h}} - \int_{\R + 2\pi \im (\tilde{h}+1)} +  \int_{\R + 2\pi \im \tilde{h}}    -  \int_{\R + 2\pi \im (\tilde{h} - 1)}        \right)   e^{-S_{\text{eff}}(\phid)} \mathrm{d} \phid \,. 
\end{equation}
\begin{figure}
    \centering
    \includegraphics[width=\textwidth]{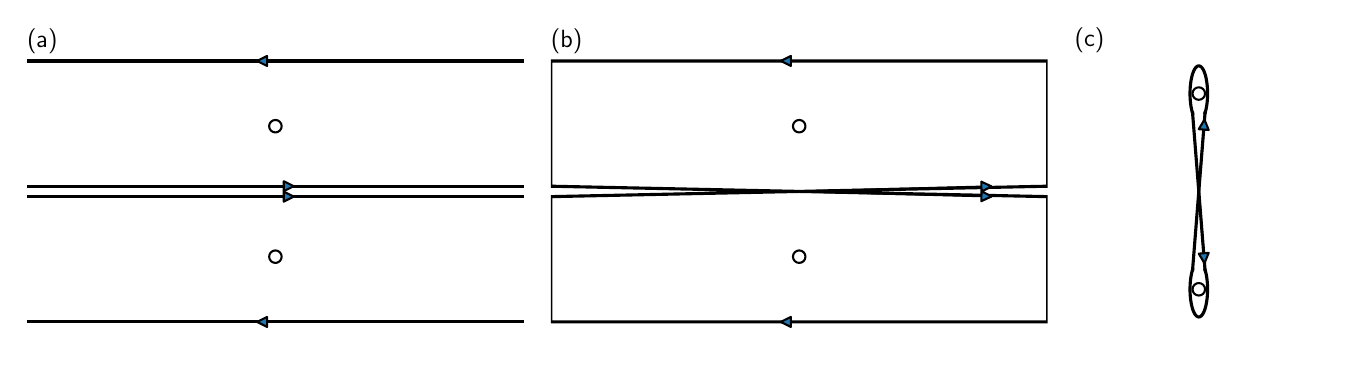}
    \caption{The deformation of contour in \eqref{eq:ZU1U2-contour}. Panel (a) represents the original contour in \eqref{eq:ZU1U2-contour}. In Panel (b), we connect the infinities of (a) in a way compatible with the branch choices (the empty circles represents the two relevant branch cut singularities). We obtain a figure ``8'', which is further deformed into Panel (c), which can be written as the difference between two integrals on the interval connecting the singularities. }
    \label{fig:contours}
\end{figure}
The contour of $\phid$ is depicted in Figure~\ref{fig:contours} (a), and the branch of the logarithm in the effective action
$$ S_{\text{eff}}(\phid) = \frac{\phid^2}{2b^2} + \frac{2\eta}{b^2}\ln \left(2 \cosh\left( \frac{\phid}2 \right) \right) 
$$
is chosen such that $- \Im ( \ln \cosh (\phid/2) )/ 2\pi = h_0, h_0 + \frac12, h_0 + 1 , h_0 + \frac12 $ for the four terms, respectively. Therefore, we can deform the contour into the figure ``8'' depicted in Figure~\ref{fig:contours}-(b) and (c)~\footnote{This contour is a simpler version of the Pochhammer contour used to define the Beta functions.}. Indeed, the way the contour winds around the two neighboring zeros of the $\cosh$ (depicted as empty dots in the Figure) is precisely compatible with the branch choices. Now, we observe that the contour of Figure~\ref{fig:contours} (c) is precisely the sum of two elementary cycles of the effective action:
\begin{align}
       Z(\mathcal{U}_{h_1, h_2}) =& \Gamma(2\eta / b^2) \left( \widetilde{Z}(\mathcal{I}_{\tilde{h}}^{h_0}) - 
      \widetilde{Z}(\mathcal{I}_{\tilde{h}}^{h_0 + 1  })\right) \nonumber \\
      =&\Gamma(2\eta / b^2)  \widetilde{Z}(\mathcal{I}_{h_1,h_2}) (1 - e^{ 4\pi \im \eta/b^2 }) \label{eq:A5} 
\end{align}
where in the second line we used the notation \eqref{eq:Idef2} and the fact that the effective action on $ \mathcal{I}_{\tilde{h}}^{h_0 + 1 }$ and that on $\mathcal{I}_{\tilde{h}}^{h_0}$ plus $-2 \pi \im \times 2\eta / b^2$. Eq.~\eqref{eq:A5} is equivalent to \eqref{eq:ZandZCG} by the reflection formula.

 \section{Thimble numerics}\label{sec:num}
 In this appendix we describe how to construct numerically the thimble of  an action with two degrees of freedom, $S(\varphi_1 , \varphi_2)$, $(\varphi_1, \varphi_2) \in \C^2$, associated with a critical point. Without loss of generality we can assume that the critical point is $\varphi_{1} = 0, \varphi_2 = 0$. We also assume that the Hessian is non-degenerate, that is, 
 \begin{equation}
     S(\varphi_1, \varphi_2) =  S_c + \frac12  \sum_{i,j=1}^2 \varphi_i H_{ij}  \varphi_j + \mathcal{O}(\varphi^3)  \,,\,
 \end{equation}
and $H$ is an invertible complex symmetric matrix. 

In principle, the basic idea is very simple: the steepest descent contours that end at the saddle point (as $s \to -\infty$) can be parametrized by one parameter, say the angle $\theta$ at which it approaches the saddle point. Hence, we can integrate the upward flow for a dense lattice of $\theta$'s, and obtain a good approximation of the thimble parametrized by the polar coordinate $(s, \theta)$. To integrate over the thimble, it suffices to know the Jacobian $\partial(\varphi_1, \varphi_2) / \partial(s, \theta)$; indeed, 
\begin{equation}\label{eq:B2}
    \int_{\mathcal{T}} e^{-S} \mathrm{d} \varphi_1 \mathrm{d} \varphi_2 =  \int_{0}^{2\pi} \mathrm{d}\theta\int_{-\infty}^\infty \mathrm{d} s  \,  e^{-S (\varphi_1, \varphi_2)} \,   \det \left[ \frac{\partial(\varphi_1, \varphi_2)}{\partial(s, \theta)} \right] \,.
\end{equation}
The Jacobian can be obtained by integrating the ``adjoint flow''~\cite{Alexandru:2020wrj}, see \eqref{eq:adjoint} below.

There is however a practical issue which renders the above naive approach numerically unfeasible. The upward flow, linearized near the saddle point:
\begin{equation}
    \partial_s \varphi_i = \sum_j \overline{H_{ij} \varphi_j} + \mathcal{O}(\varphi^2)
\end{equation}
will have two different local Lyapunov exponents. As a result, the polar coordinate is a singular parametrization of the thimble, which is problematic for numerical integration. To resolve this issue, we shall find a linear change of variables 
\begin{equation}
    \begin{pmatrix} \varphi_1 \\ \varphi_2 \end{pmatrix} = U \begin{pmatrix} \phi_1 \\ \phi_2 \end{pmatrix}
\end{equation}
such that
\begin{equation}\label{eq:B5}
    U^{T} H U = I \,.
\end{equation}
Then, the upward flow in the coordinate system $\phi$, 
\begin{equation} \label{eq:flowphi}
    \partial_s \phi_i = \overline{\frac{\partial S}{\partial \phi_i} } \,,
\end{equation} 
will have a linearization with equal Lyapunov exponents. Now, the upward flow is \textit{not} invariant under a change of variable. Rather,  \eqref{eq:flowphi} transforms to 
\begin{equation} \label{eq:flowright}
    \partial_s \varphi_i = g_{ij} \overline{\frac{\partial S}{\partial \varphi_j} } \,,\, \text{where } g = {U} U^\dagger \,.
\end{equation}
Note that $g$ is an invertible Hermitian matrix, and thus can be viewed as a metric. The modified flow \eqref{eq:flowright} will change the geometric shape of the thimble, but not its topology, so the path integral will have the same value. 

We now detail the numerical workflow, given the action $S(\varphi_1, \varphi_2)$ and a non-degenerate saddle point, assumed to be at $\varphi_1 = \varphi_2 = 0$. 
\begin{enumerate}
    \item Compute the Hessian $H_{ij} = \partial^2 S / ( \partial \varphi_i \partial \varphi_j)$ at the saddle point, and find the matrix $U$ as in \eqref{eq:B5}. This can be done as follows. We diagonalize $H$ so that $v^{-1} H v = w$, where $w$ is a diagonal matrix. The symmetry of $H$ ensures that $D = v^{T} v$ is also a diagonal matrix (or can be made so in case of degeneracy). Then we let $U = (v^{-1})^T D^{\frac12} w^{-\frac12}$ (the branch choice of the square roots can be arbitrary; we assume that the saddle point is non-degenerate so $w$ is invertible). 
    \item Choose a mesh of $\theta \in [0, 2\pi]$ (a uniform grid of $10^2$ points is sufficiently accurate). For each $\theta$ in the mesh, carry out step 3 and 4.
    \item  Solve the upward flow~\eqref{eq:flowright} with initial condition $$ \begin{pmatrix} \varphi_1 \\ \varphi_2 \end{pmatrix}_{s=0} = \epsilon \, U \begin{pmatrix} \cos(\theta) \\ \sin(\theta) \end{pmatrix} \,, $$
    where $\epsilon$ is a small number, e.g, $\epsilon = 0.05$. We stop the the flow at  $s = s_{\max}$; the truncation is triggered by the action exceeding the saddle-point value $S_c$ plus a large difference, e.g., $20$.
    \item Compute the Jacobian matrix. The derivatives $\partial_s \varphi_i$ are obtained by differentiating the flow solution just obtained. The $\partial_s$ derivatives are obtained by integrating the adjoint flow equation:
    \begin{equation}\label{eq:adjoint}
         \partial_s \, \frac{\partial \varphi_i}{\partial \theta} = g_{ij} \overline{\left(\frac{\partial^2 S}{\partial \phi_j \partial \varphi_k} \right) }  \frac{\partial \varphi_k}{\partial \theta} \,,\, \left.\frac{\partial \varphi_i}{\partial \theta}\right|_{s=0} = \epsilon U  
         \begin{pmatrix} - \sin(\theta) \\ \cos(\theta) \,. \end{pmatrix} 
    \end{equation}
   \item Calculate the integral \eqref{eq:B2} numerically (the $s$-integral's limit is from $0$ to $s_{\max}$.) In practice, we perform the $s$-integral for each $\theta$ in the mesh, interpolate the results by a smooth function of $\theta$, and integrate the latter. Finally, we add the contribution of the disk $B = \{\phi_1^2 + \phi_2^2 < \epsilon^2 \}$ (see step 3 above), computed using a Gaussian approximation of the action in that disk:
   $$ \int_{B} e^{-S} \mathrm{d} \varphi_1   \mathrm{d} \varphi_2 \approx  2\pi \det(U) (1 - e^{-\epsilon^2/2}) e^{-S_c} $$
\end{enumerate}

	\bibliographystyle{JHEP}
	\bibliography{ref}
	
\end{document}